\documentclass{svjour3}                     
\smartqed  
\usepackage{graphicx}
\usepackage{psfrag}
\usepackage{epsfig}
\usepackage{amsmath}
\usepackage{color}
\journalname{Experiments in Fluids}
\begin{document}

\title{Estimation of burst-mode LDA power spectra
}


\author{Clara M. Velte         \and
        William K. George  \and
        Preben Buchhave
}


\institute{Clara M. Velte \at
            Department of Mechanical Engineering\\
            Technical University of Denmark\\
              Nils Koppels All\'{e} Bldg. 403 \\
              2800 Kgs. Lyngby, Denmark\\
              Tel.: +45-45254342\\
              Fax: +45-45930663\\
              \email{cmve@dtu.dk}
             \and
           William K. George \at
           Department of Mechanical and Aerospace Engineering\\
           Princeton University\\
           Princeton, NJ 08544\\
            and\\
            Department of Aeronautics\\
            Imperial College London\\
            South Kensington Campus\\
            London SW7 2AZ\\
           \and
           Preben Buchhave \at
            Intarsia Optics\\
            S{\o}nderskovvej 3\\
            3460 Birker{\o}d, Denmark\\
}

\date{Received: date / Accepted: date}

\maketitle

\begin{abstract}
The estimation of power spectra from LDA data provides signal processing challenges for fluid dynamicists for several reasons: acquisition is dictated by randomly arriving particles, the registered particle velocities tend to be biased towards higher values and the signal is highly intermittent. The signal can be interpreted correctly by applying residence time weighting to all statistics and using the residence-time-weighted discrete Fourier transform to compute the Fourier transform. A new spectral algorithm using the latter is applied to two experiments; a cylinder wake and an axisymmetric turbulent jet. These are compared to corresponding hot-wire spectra as well as to alternative algorithms for LDA signals such as the time slot correlation method, sample-and-hold and common weighting schemes. 
\end{abstract}

\section{Introduction}
\label{intro} The burst-mode LDA should be operated with at most one particle in the scattering volume, which means that for most of the time there are none~\cite{1,2,Deadtime1,3,5,6}. Thus the burst-mode LDA can be assumed to sample whenever a particle passes through the scattering volume. This presents four challenges to interpreting the signal correctly:\\

\noindent \textit{Random (but velocity dependent) sampling:} The particle arrivals dictate the sampling which is therefore non-uniform and velocity dependent. In other words, the sampling process and sampled process are not statistically independent.\\

\noindent \textit{Velocity bias:} The statistics can be significantly biased since the sampling process itself depends on the velocity. Since the probability of acquiring more samples per unit time increases with higher velocities, the bias is usually (but not always) towards higher velocities~\cite{2}.\\

\begin{figure}
\includegraphics{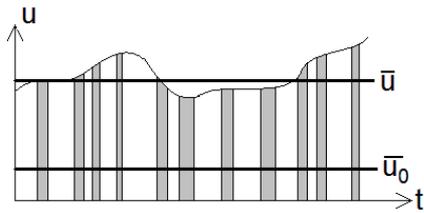}
\caption{Sketch of velocity, u (curve), along with the interrupted
sampled signal, u$_0$ (shaded areas), resulting from random arrivals
of particles carried by the flow.} \label{fig:1}
\end{figure}

\noindent \textit{Intermittent signal:} The LDA samples only when a particle traverses through the measuring volume, meaning that most of the time there is no signal present. Figure~\ref{fig:1} illustrates this where the continuous curve represents a velocity. This velocity is sampled by an LDA, where the sampled signal is represented by the grey bars. The on-off nature of the sampled LDA signal changes dramatically the power spectrum if it is processed as an analog signal; and it produces incorrect moments if simply processed using arithmetic algorithms developed for data sampled at a fixed rate.\\

\noindent \textit{Dead time:} Like all practical instruments the burst mode LDA must average over some finite time (and volume) during which it is not able to acquire new data due to the restriction of measuring at most one particle at a time. Further, the processor may require a finite time for data transfer etc. during which the processor again cannot acquire new data. These two effects effectively lock out the acquisition of new samples during some finite dead time $\Delta t_d$. For a measurement technique such as the LDA where the sampling is random, this limitation in the acquisition can sometimes have a dramatic adverse effect on the resulting power spectrum, significantly affecting the energy across the whole range of frequencies~\cite{Deadtime1}.\\

For the signal to be interpreted correctly and to avoid velocity bias, one must apply residence time-weighting to all statistical analysis~\cite{3,5,6} (see also Appendix A for a discussion on velocity and directional bias). This claim can be justified singlehandedly by the rigorous theory presented in section~\ref{sec:1}, but is also further supported by the results obtained from the measurements in section~\ref{sec:5results}. Note that the only assumption in this theory is that the seeding is randomly distributed initially. In other words, we are not making any unphysical assumptions about the flow (e.g., multivariate Poisson, joint normality etc.). Note also that residence time weighting is not subject to directional bias caused by the variation of the geometrical cross section as viewed from different flow directions when the measuring volume is not spherical (see Appendix A).

We will first review the theoretical basis for residence time weighting. Then the theory will be used to develop a practical algorithm for the estimation of the power spectrum from burst-mode LDA data. Some flaws in the earlier theory~\cite{1,3}, the goal of which was to provide an unbiased and unaliased spectral estimator from the random samples, have been identified and corrected. Further, a stronger theory is presented with more relaxed conditions regarding the ability of the particles to follow the flow. The new methodology is illustrated using experiments in an axisymmetric turbulent far jet and a cylinder wake with a statistical basis and accuracy previously not remotely available due to the lack of computing power at the time the theory was developed in the 1970's~\cite{1,3,5}. The results are compared to corresponding hot-wire measurements as well as the time slot correlation method and spectra from interpolated and re-sampled (sample-and-hold) LDA signals. Further, alternative weighting schemes are tested and evaluated for completeness. Due to the random sampling, the variability of the spectral estimator is larger than for regularly sampled data, so this will also be discussed. Finally, a Matlab script for fast computation of burst-mode LDA spectra using the direct method is provided in Appendix B.

\section{Residence time weighting}
\label{sec:1}

\subsection{First order statistics}
\label{sec:1.1}

The problem of representing the burst-mode LDA signal has previously been addressed by considering the fluid particle motions, $\vec{x}(\vec{a},t)$, in Lagrangian space~\cite{3,5,6,7}, and using a sampling function, $g(\vec{a})$, that samples the velocity at the initial spatial locations of the particle. Subsequent to this it was assumed that the flow was incompressible and that the sampling function could be transformed to Eulerian space. This part of the analysis appears to have been incorrect. Also the earlier analysis did not include the possible effects of the particles not following the flow. This is especially important since clustering of particles has been observed in recent years~\cite{8}; therefore this part of the analysis has also been reconsidered.

Considering the particle motions in Lagrangian space, the velocity as sampled by the LDA at location $\vec{x}$, say $\tilde{u}_{i0}(\vec{x},t)$, can be defined by
\begin{equation}
\tilde{u}_{i0}(\vec{x},t) =
\int\!\!\!\int\!\!\!\int_{\mathrm{all\;space}}u_i(\vec{a},t)
g(\vec{a}) w \left ( \vec{x} - \vec{X}[\vec{a},t] \right ) \,
d^3\vec{a}
\end{equation}
where $u_i(\vec{a},t)$ is the velocity of the i'th \textit{particle} with initial position $\vec{a}$. $g(\vec{a})$ is a sampling function that describes whether a particle is present or not at position $\vec{a}$ at the arbitrarily chosen initial instant. The positions of the particles are given by their displacement field, say $\vec{X}[\vec{a},t]$. $w \left ( \vec{x} \right )$ is a weighting function that accounts for the finite extent of the measuring volume centered at location $\vec{x}$, and effectively `turns on' when the particle enters the volume and `turns off' when it leaves. The weighting function is dimensionless and must by definition yield the volume of the measuring volume when integrating across all space, i.e.,
\begin{equation}
V(\vec{x}) \equiv \int\!\!\!\int\!\!\!\int_{\mathrm{all\;space}} w
\left ( \vec{x}' \right ) \, d^3\vec{x}'
\end{equation}

It has long been recognized that the intermittently sampled velocity (illustrated in Figure~\ref{fig:1}) yields biased and incorrect estimates of the moments such as the mean, \textit{\textit{unless it is treated as a time signal while it is `on'}}~\cite{3,5,6,7}. The `on-time' of the signal is given by
\begin{equation}
\int\!\!\!\int\!\!\!\int_{\mathrm{all\;space}} g(\vec{a}) w \left (
\vec{x} - \vec{X}[\vec{a},t] \right ) \, d^3\vec{a} = \mu V \approx
\frac{1}{T} \sum_{n=0}^{N-1} \Delta t_n
\end{equation}
where $\mu = \overline{g \left ( \vec{a} \right )}$ is the expected number of particles per unit volume, $\Delta t_n$ is the time spent in the volume by the $n$-particle, $T$ is the record length and $V$ is the volume of the scattering volume. Hence, $\mu V$ is the expected number of particles in the volume, but is also the percentage of the total time for which a particle is present.

As noted above, the remedy to the problem (for a statistically stationary process) portrayed in Figure~\ref{fig:1} is to account for the time that the signal is actually present. Thus the mean of the burst-mode signal is the mean of the volume-averaged velocity, but multiplied by $\mu V$ (which is always less than unity); i.e.:
\begin{eqnarray}
\overline{\tilde{u}_{i0}} & = &
\overline{\int\!\!\!\int\!\!\!\int_{\mathrm{all\;space}}u_i(\vec{a},t)
g(\vec{a}) w \left ( \vec{x} - \vec{X}[\vec{a},t] \right ) \,
d^3\vec{a}} \\\nonumber 
& = & \mu V \bar{u}_i \\\nonumber & \approx & \lim_{\substack{T
\rightarrow \infty \\ N \rightarrow \infty}} \frac{1}{T} \left \{
\sum_{n=0}^{N-1} \int_{t_n}^{t_n+\Delta t_n} u_{i0} (t)\, dt \right
\}, \nonumber
\end{eqnarray}
where the time integral can be used to replace the ensemble average if the process is stationary. From this it is straightforward to see that the piecewise mean velocity integral can be best approximated by:
\begin{equation}
\bar{u}_i = \frac{\sum_{n=0}^{N-1} u_i(t_n)\Delta
t_n}{\sum_{n=0}^{N-1} \Delta t_n}
\end{equation}
where $\Delta t_n$ is the residence (or transit) time for the n$^\mathrm{th}$ realization. Also it is clear at this point precisely which velocity, $\bar{u}_i$, is being measured: namely the \textit{volume-averaged particle velocity}. Only if the particles are following the flow does this correspond to the volume-averaged Eulerian velocity.

Similar residence time weighted algorithms can be derived for all the single-time moments, and must be used when correlating burst-mode LDA data with continuous signals as well (like from hot-wires or pressure sensors), c.f.~\cite{1,9}. The estimator of the variance can be obtained in the same manner as the estimator of the mean value, yielding
\begin{equation}\label{eq:var}
\overline{u^2_i} = \frac{\sum_{n=0}^{N-1} \left [ u_i(t_n) - \bar{u}_i \right]^2 \Delta
t_n}{\sum_{n=0}^{N-1} \Delta t_n}
\end{equation}

Though this seems to have been forgotten, the residence time weighting is and has long been recognized to be the in-principle correct way to weight \textit{\textbf{all}} statistics (including the discrete Fourier transform), assuming only that the particle seeding is spatially homogeneous (see, e.g., Albrecht \textit{et al.}~\cite{2} pp. 552, \cite{6a}). Other weighting schemes have sometimes been chosen mainly because the available commercial signal processors have not always provided reliable values for the residence times. Though not all issues seem to have been solved (see section~\ref{sec:7} for a discussion), this does not seem as problematical today as it has been previously. Due to these practical shortcomings, numerous alternative methods have been proposed (and are also tested in section~\ref{sec:AWEs}) with reference to practical problems with the measurement of the residence time. These are however generally not recommended~\cite{2}, the reason being apparent from the results in section~\ref{sec:AWEs} of this paper.

As noted above and by~\cite{6}, the residence time should also be used to obtain the Fourier coefficients for spectral estimation; i.e., discretizing the finite time Fourier transform of the velocity signal yields:
\begin{equation}\label{eq:DFT}
\hat{u}_{i\,T} (f)= \int_{0}^{T}e^{- i 2 \pi ft} u_i(t)\, dt \approx
\sum_{n=0}^{N-1} e^{- i 2 \pi ft_n} u_{i\,n} \Delta t_n
\end{equation}
where $u_{in} = u_i(t_n)$, and the subscript $T$ indicate that the quantity is evaluated over a finite time domain. The spectral estimator which can be derived from this is discussed in detail in section~\ref{sec:3} below, following a section of the theoretical implications of the intermittency of the burst mode LDA output interpreted as a time signal. Note that the residence-time Fourier transform and spectra computed from it in this manner are {\it not} aliased, so the usual anti-aliasing measures for equally sampled data (whether using instantaneous data or time-slotted correlations) are not necessary.

\subsection{Second order statistics} \label{sec:2}

The two-time cross-correlation of the instantaneous burst-mode LDA signal is given by:
\begin{equation}\label{eq:sost}
\overline{\tilde{u}_{i0}(t)\tilde{u}_{j0}(t')}=
\mathop{\int\!\!\!\int\!\!\!\int\!\!\!\int\!\!\!\int\!\!\!\int}_{\quad
\mathrm{all\;space}}\overline{u_i(\vec{a},t) w \left (
\vec{x}[\vec{a},t] \right ) u_j(\vec{a}',t') w \left (
\vec{x}[\vec{a}',t'] \right )} \,\, \overline{g(\vec{a})
g(\vec{a}')} \, d^3\vec{a} \,d^3\vec{a}'
\end{equation}

Since the initial positions of the particles (the $\vec{a}$'s) are truly independent of the field that transports them, the statistics of the $g$'s and velocity field can be assumed to be statistically independent\footnote{Note that this uncoupling is crucial, since otherwise the sampling process and sampled process are not statistically independent, which means the statistics are in principle biased. Note also that no arbitrary assumptions (e.g. Poisson-distributed, etc.) are necessary to derive the residence-time weighted algorithms, quite unlike competing bias correction schemes~\cite{9b}. In effect, by this residence-time-weighting approach the bias is simply avoided.}. From before we already know that for the sampling function (cf.,~\cite{3,5})
\begin{equation}
\overline{g(\vec{a}) g(\vec{a}')} = \mu^2 + \mu \delta \left (
\vec{a}' - \vec{a} \right )
\end{equation}
This yields
\begin{eqnarray}\label{eq:2pointstat}
\lefteqn{\overline{\tilde{u}_{i0}(t)\tilde{u}_{j0}(t')}= }
\\\nonumber & & \mu^2
\int\!\!\!\int\!\!\!\int\!\!\!\int\!\!\!\int\!\!\!\int_{\mathrm{all\;space}}\overline{\left
[ u_i(\vec{a},t) w \left ( \vec{x}[\vec{a},t] \right ) \right ]
\left [ u_j(\vec{a}',t') w \left ( \vec{x}[\vec{a}',t'] \right
)\right ]} \, d^3\vec{a} \,d^3\vec{a}' +\\\nonumber & & \mu
\int\!\!\!\int\!\!\!\int_{\mathrm{all\;space}}\overline{\left [
u_i(\vec{a},t) w \left ( \vec{x}[\vec{a},t] \right ) \right ] \left
[ u_j(\vec{a},t') w \left ( \vec{x}[\vec{a},t'] \right )\right ]} \,
d^3\vec{a} \nonumber
\end{eqnarray}
The double integral on the RHS is just the cross-correlation of the \textit{instantaneous volume-averaged particle velocity at two times}; i.e.,
\begin{equation}
\mu^2
\mathop{\int\!\!\!\int\!\!\!\int\!\!\!\int\!\!\!\int\!\!\!\int}_{\quad
\mathrm{all\;space}}\overline{\left [ u_i(\vec{a},t) w \left (
\vec{x}[\vec{a},t] \right ) \right ] \left [ u_j(\vec{a}',t') w
\left ( \vec{x}[\vec{a}',t'] \right )\right ]} \, d^3\vec{a}
\,d^3\vec{a}' = (\mu V)^2
\overline{\tilde{u}_{i0}^{\mathrm{vol}}(t)\tilde{u}_{j0}^{\mathrm{vol}}(t')}
\end{equation}
As above, it corresponds to the two-time cross-correlation of the Eulerian velocities \textit{only} if the scattering particles are following the flow.

The second integral is a consequence of the intermittency of the characteristic LDA off-on signal; see Chapter 10 in~\cite{4}. The quantity under the integral sign is evaluated for only a single particle, since it requires the initial particle positions to be the same ($\vec{a}' = \vec{a}$) and the particles to spend only a very short time in the scattering volume so that $t' \approx t$. Thus to the level of our measuring equipment, the entire expression can be approximated by:
\begin{equation}
\mu \int\!\!\!\int\!\!\!\int_{\mathrm{all\;space}}\overline{\left [
u_i(\vec{a},t) w \left ( \vec{x}[\vec{a},t] \right ) \right ] \left
[ u_j(\vec{a},t') w \left ( \vec{x}[\vec{a},t'] \right )\right ]} \,
d^3\vec{a} \approx \mu V \left [
\overline{\tilde{u}_{i0}^{\mathrm{vol}}(t)\tilde{u}_{j0}^{\mathrm{vol}}(t')}\right
]
\end{equation}
where we have assumed that $\left [ w (\vec{x}) \right ]^2 \approx w (\vec{x})$, since $w$ is usually just `1' or `0' (This is why the integral produces a V, not a V$^2$). Note that the quantity $\left [\overline{\tilde{u}_{i0}^{\mathrm{vol}}(t)\tilde{u}_{j0}^{\mathrm{vol}}(t')} \right ]$ still includes both the mean and fluctuating components, so it can be very large indeed. The reason why this term becomes so large is just a consequence of subtracting off the wrong mean, since the mean value of the signal, $\bar{u}$, and the interrupted signal mean, $\bar{u}_0$, typically differ considerably, see Figure~\ref{fig:1}. Thus expression (\ref{eq:2pointstat}) simplifies to:
\begin{equation}\label{eq:2pointstat2}
\overline{\tilde{u}_{i0}(t)\tilde{u}_{j0}(t')} \approx (\mu V)^2
\overline{\tilde{u}_{i0}^{\mathrm{vol}}(t)\tilde{u}_{j0}^{\mathrm{vol}}(t')}
+ \mu V \overline{\left [
\tilde{u}_{i0}^{\mathrm{vol}}(t)\tilde{u}_{j0}^{\mathrm{vol}}(t')
\right ] \delta (t'-t)}.
\end{equation}

The second term on the RHS contains the added variance and the constant spectral offset arising from the random arrivals which is uncorrelated with itself at all but zero time-lag where the correlation is perfect. This term is a consequence of the intermittency of the randomly sampled signal. As first noted by~\cite{7} (see also~\cite{3,5}), the second term produces a spike in the origin of the autocovariance or, respectively, a constant offset in the spectrum. It should be obvious that the effect of this term must be removed, for example, when computing the autocovariance directly from the LDA data using the time-slot technique (see~\cite{3}). The correlation there can only be obtained by extrapolating the rest of the curve to $\tau=0$ (using the approximation of parabolic autocovariance near $\tau=0$) or, correspondingly, by removing the high-frequency asymptote (constant spectral offset) in the spectrum~\cite{4}. Both techniques should be applied with caution since they have inherent limitations due to the way the data is reduced in the acquisition process. The former due to time resolution of the slots and filtration, which causes an added contribution around the origin in the autocovariance and hence a Taylor microscale which is determined by the instrument filtration and not by the flow~\cite{GL1973}. The latter technique is affected by effects of windowing and dead time~\cite{Deadtime1} (see section~\ref{sec:deadtime} for a brief description), but may still work well in practice for a well-designed experiment.

\section{Practical algorithms}

\subsection{Direct spectral estimator} \label{sec:3}

The spectral estimator (originally derived by~\cite{3,6} and corrected by~\cite{4}) can be written in short form as:
\begin{equation}
S_0(f) = \frac{1}{(\mu V)^2} S_{0_T}^{B}(f) = \frac{T^2}{\left (
\sum_{n=0}^{N-1} \Delta t_n \right )^2 } \frac{\left |
\hat{u}_{i\,T}(f) \right | ^2}{T} \label{eq:directestimatorwkg}
\end{equation}
The factor $1/(\mu V)^2 = T^2 / \left ( \sum_{n=0}^{N-1}\Delta t_n \right )^2$ compensates for the fact that there is signal only when a particle is in the measuring volume (as shown in equation~\ref{eq:2pointstat2}). Like all spectral estimators (and as shown later when the variability is considered), this one too must be either smoothed or block-averaged to produce converging statistics. If the latter, it is essential that all the blocks be of the same length in time.

Substituting for the discretized Fourier transform using equation (\ref{eq:DFT}) yields:
\begin{equation}\label{eq:specest}
S_0(f) = \frac{T}{\left ( \sum_{n=0}^{N-1} \Delta t_n \right )^2 }
\mathop{\sum_{m=0}^{N-1} \sum_{n=0}^{N-1}}_{m \neq n} e^{- i 2 \pi f
(t_m - t_n)} u_m u_n \Delta t_m \Delta t_n
\end{equation}
which is exactly the estimator originally employed by~\cite{3}. Note that the self-products are omitted to eliminate the systematic error contained in the last term of equation~(\ref{eq:2pointstat2}). Also note that the frequencies can be chosen arbitrarily since the Fourier transform is computed using the Discrete-Time Fourier Transform (DTFT). In this technique logarithmically spaced estimates (or any other system for frequency selection) are easily implemented.\footnote{Note that due to the finite LDA measuring volume, it does not make sense to evaluate the spectra above the probe-volume-cut-off frequency, say $f_c$, and  expect reliable results (this is, of course, true regardless of algorithm). Typically $f_c \sim U_c/d$ where $d$ is a length characterizing the volume and $U_c$ is a convection velocity. Some techniques, such as interpolation and resampling, can provide continuous and smooth spectra even far beyond the measuring volume cut-off frequency.  These are of course artifacts and not at all representative of the actual physics.} George~\cite{6} suggested that the computations could be significantly reduced by computing first the direct transform of equation (\ref{eq:DFT}) and then squaring the result (and subtracting the diagonal $m=n$ trace). This produced the same result with more efficiency especially for longer records and has been implemented in Appendix B.

There is a distinct difference between the sampling methods that are most straightforwardly used to form the spectral estimators based on the autocovariance and the direct estimator (\ref{eq:specest}), a difference, which is further discussed in section 6.2 of~\cite{3}. The autocovariance function is formed from batches of a certain number of data points. In the direct estimator it is most practical to form the estimate by the method of block averaging (analogous to the method described in Welsh 1967). Each block or batch estimate is formed from an expression like eq. (\ref{eq:specest}), but like all practical signals including a spectral window, which defines the bandwidth of the estimate and determines the end of the summation. Because of the random sampling the direct estimator is formed from a random number of samples. This is not a problem.

Since \textit{all} spectra are windowed by the window imposed on it by the length of each block of data, each data block must be much longer than the integral scale to reduce spectral leakage.  Even so additional windows can be imposed to reduce leakage even further than the $1-|\tau|/T$ window which naturally results from the direct transform method.  Note that in this regard, sampling for a fixed period (resulting in a random number of samples in a batch) with the residence time weighting applied is completely equivalent to the conventional equispaced sampling with fixed batch length.

\subsubsection{Statistical convergence and sample size} \label{sec:6}

The random sampling inherent in the burst-mode LDA measurement
technique significantly affects the amount of data required to
obtain converged burst-mode LDA spectra. If one relies on the
assumption that the fourth order moments are jointly Gaussian, i.e.,
$\overline{uu'u''u'''} = \overline{uu'}\,\,\overline{u''u'''} +
\overline{uu''}\,\,\overline{u'u'''} +
\overline{uu'''}\,\,\overline{u'u''}$, then the variability of the
spectral estimator for a single block of data can be shown to be
given by (see Appendix D in~\cite{4});
\begin{equation}\label{eq:converg}
\varepsilon_S^2 = \frac{\mathrm{var}\{S(f)\}}{\left [ S(f) \right
]^2} = 1 + \frac{4}{\nu}\frac{B(0)}{S(f)}
\end{equation}
where $B(0)$ is the velocity variance. When many blocks of data are averaged
together, this variability decreases by the inverse square root of
number of independent data records used to produce the
block-averaged spectrum. Clearly when $S(f)$ drops (as it usually does
at higher frequencies), the variability increases. Or said another
way, the number of blocks of data required to achieve a given
variability of the spectral estimator increases with decreased
magnitude of the spectrum~\cite{1,14}. This is quite different from
the usual spectra (i.e., not randomly sampled), for which the
variability is frequency independent.

\subsubsection{The impact of dead-time} \label{sec:deadtime}

In principle, random sampling eliminates aliasing in the case of ideal sampling where the sampling function can be represented by a sharp delta function. However the spatial and temporal resolution of the acquired data are always finite for one reason or another, and ultimately limited by the finite size of the measuring volume. This is true for all acquired signals since \textit{no instrument can measure instantaneously, but must always average over some finite time/space}. Since the burst-mode LDA can only measure one particle at a time, each measurement is associated with a dead-time corresponding to the time it takes for the particle to pass through the measuring volume (the residence time). This dead time may be further extended by additional processing time etc.. Further, the measurement is also averaged and filtered. This will affect the values around the origin in the autocovariance and naturally also therefore the corresponding spectrum.

If one considers the simple case of point sampling, where averaging across the processing time is not considered, it can be shown that the effect of a constant dead-time $\Delta t_d$ on a randomly sampled signal with at most one sample in the measuring volume at a time can be represented by (see~\cite{Deadtime1})
\begin{equation}
S_{0,\Delta t_d}(f,\Delta t_d) = \frac{\overline{u^2_{0,\Delta t_d}}}{\nu_0} + S(f) \otimes \left [ \delta (f) - 2\Delta t_d \, \mathrm{sinc} (2 \pi f \Delta t_d) \right ] \label{eq:ucontrbdt}
\end{equation}
where $S(f)$ is the true spectrum and $\nu_0$ is the reduced sample rate due to dead time, $\nu_0 = \nu e^{-\nu \Delta t_d}$. The first term is a constant offset which is only a consequence of how the spectrum is calculated. The effects of the sinc term vanish in the limit of $f \rightarrow \infty$ and/or for $\Delta t_d = 0$.

In practice, the situation for the LDA is even somewhat more complicated, involving variable (but measurable) residence times that are correlated with the sampling process (i.e., the flow bringing the particles).

\begin{figure*}
\begin{minipage}{0.5\linewidth}
  \center{\includegraphics[width=0.95\textwidth]{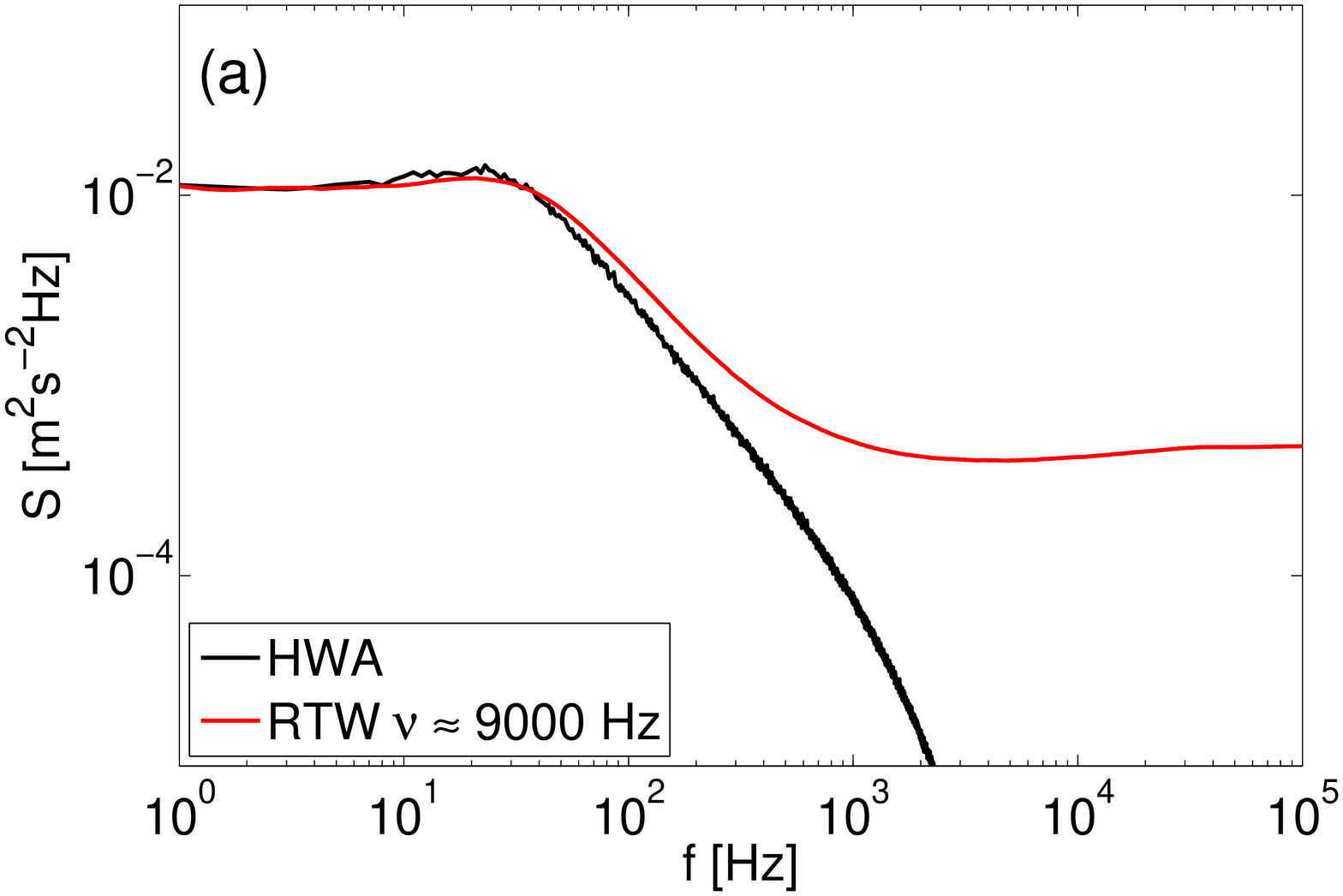}}
\end{minipage}\vspace{0.5cm}
\begin{minipage}{0.5\linewidth}
  \center{\includegraphics[width=0.95\textwidth]{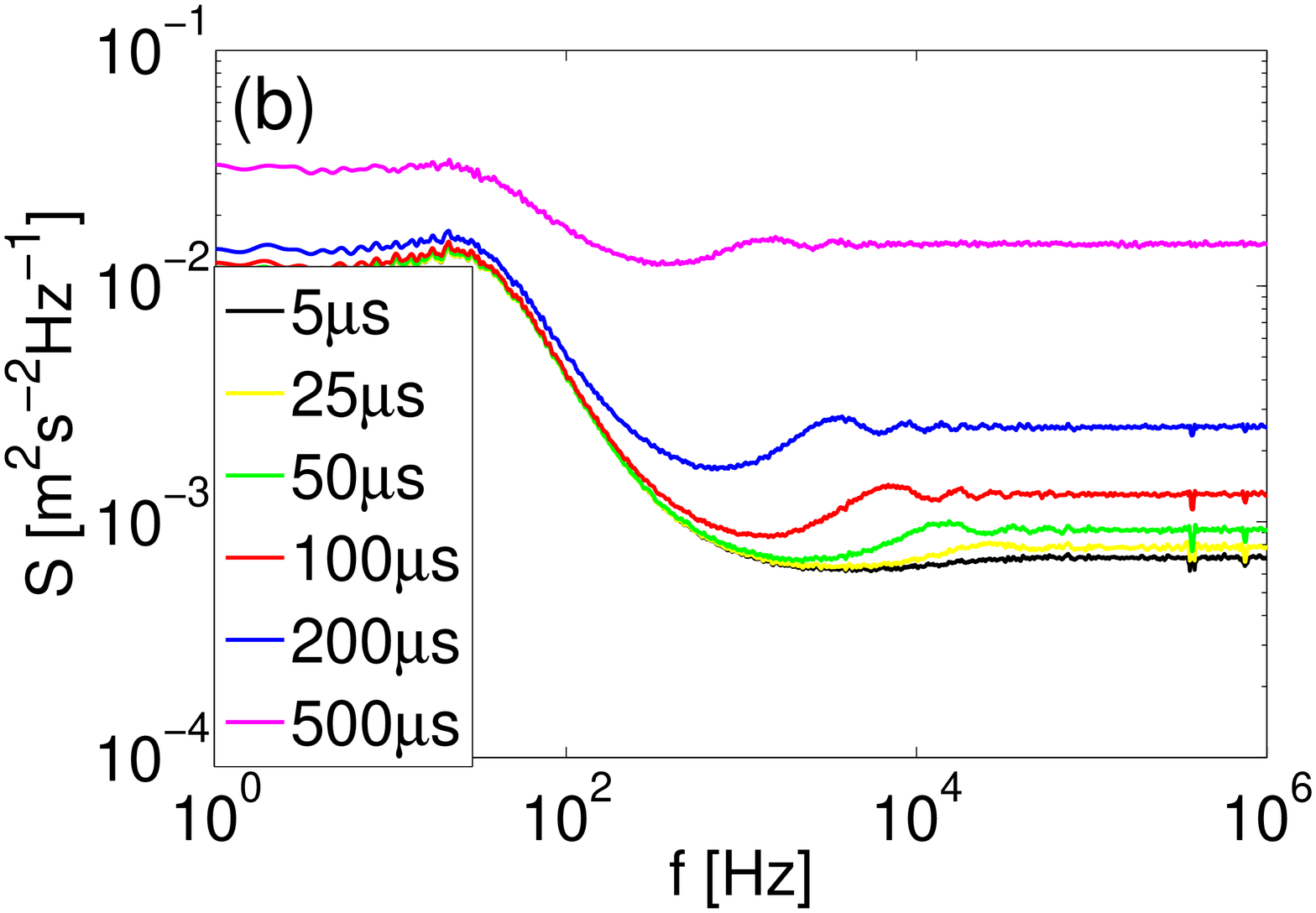}}
\end{minipage}
\caption{Measured jet power spectra obtained (a) for varying residence times as originally provided by the LDA measurements ($\nu \approx 9000$\,Hz) and (b) for $\nu = 6\,400$\,Hz with imposed constant dead-times.} \label{fig:deadt}
\end{figure*}

Figures~\ref{fig:deadt}(a) and (b) show examples of measured power spectra for the jet displaying the effects of the finite dead-time. The first figure shows how the spectrum is affected by varying residence times as originally provided by the LDA measurements. The second figure shows a burst-mode LDA spectrum with a selection of artificially imposed constant dead-times. The dead-times were introduced by excluding any measurement that succeeded another one within the prescribed dead-time. The sinc-pattern, predicted by equation~(\ref{eq:ucontrbdt}), is seen to shift to lower frequencies with increased dead-times, as expected. It is clear from the figures that the dip is more pronounced when imposing a constant dead time, which is hardly surprising since the effects are more concentrated to a single sinc-function rather than being smeared out across a distribution of occurring residence times. A deeper analysis of the dead time problem can be found in~\cite{Deadtime1}.

Note that, since this is an inherent property of the acquired LDA data, \textit{all methods for computing spectra are afflicted with this problem}. Interpolation and re-sampling methods are only seemingly able to bypass these effects, since the re-sampling frequency can be set higher than the measuring volume cut-off frequency. This way it may appear as if frequencies above the measuring volume cut-off are resolved, even though it is obvious from the finite extent of the measuring volume that they can not be.

So, unfortunately, the problem of finite dead-time is \textit{always} present in LDA measurements (in fact in \textit{all} practical sampled measurements). Further, the dead-times may vary with particle velocity and path through the measuring volume. The obvious remedies are to limit the size of the measuring volume compared to the measured flow scales, which can be done in a number of ways, and speed up the processor to make it ready for new acquisitions faster. One practical solution can be to sample continuously within one burst to further reduce the dead-time (similar to the counters). Note that one then has to take care regarding phase shifts~\cite{7,GL1973}, ambiguity noise and windowing in the Fourier transform of each individual burst. The saving feature for spectral measurement is that the dead-time is of approximately the same length as both the spatial filtering cut-off of the scattering volume and also the residence time.  So its primary effect on the signal spectrum is usually on frequencies above those of interest anyway unless the dead times are very long.

\subsubsection{The current data sets} \label{sec:currentdata}

As a practical check that dead time effects do not have a significant effect on the relevant part of the spectrum, i.e., up to the probe cut-off frequency, $f_c$, the power spectrum should (as by definition) integrate to the power of the signal, i.e.:
\begin{equation}\label{eq:energy}
\int^{f_c}_{-f_c}S_0(f)\,df \approx \overline{u^2}
\end{equation}
where $\overline{u^2}$ can be checked against the variance obtained directly from the data using equation~(\ref{eq:var}). This, of course, also requires that the record length (window) be large enough relative to the integral scale (typically at least  100) so that spectral leakage does not affect the measured spectrum significantly at the frequencies of interest.\footnote{Note that spectral leakage is often misinterpreted as noise and removed in processing with filters.  One test  is whether the `filtered spectra' integrate to the variance. If not, it is usually because leakage has redistributed the energy to higher frequencies.} Note that it does not make sense to integrate beyond the measuring volume cut-off frequency, which approximately coincides with the (first) dip, since the probe cannot produce reliable measurements above this frequency. The data used in the current work have been shown to be consistent with equation~(\ref{eq:energy}) and can therefore be used reliably in the subsequent data analysis.

\subsection{Time slot method} \label{sec:2b}

Historically the approach to this problem has been through the so called time-slot correlation method. This was originally employed by Gaster and Roberts~\cite{14} for randomly sampled data where the sampling process was statistically independent of the sampled process (see George \textit{et al.} 1978~\cite{1} for a review).

Buchhave~\cite{3} was the first to correctly implement this for LDA by including the residence time weighting. The need for residence time weighting is clearly necessary from eqn. (\ref{eq:sost}), and must account for the time both particles are in the volume. The first term of (\ref{eq:2pointstat2}) represents the desired information, which in a discrete form may be written as (see~\cite{3}):
\begin{equation}\label{eq:ac}
C(\tau) = \overline{\tilde{u}_{i0}^{\mathrm{vol}}(t)\tilde{u}_{j0}^{\mathrm{vol}}(t')} \approx  \frac{1}{(\mu V)^2T} \sum^N_{i,\, j} u_0(t_i) u_0(t_j) \Delta t_{ij} (\tau)
\end{equation}
Note that this term is non-zero only when there is a particle in the volume at time $t$ \textit{and} one at $t'$. The integral should only be computed during this time of overlap, $\Delta t_{ij}$, of realizations of $u_0(t_i)$ and $u_0(t_j)$. A more thorough and practical treatment of the implementation of the time slot method can be found, e.g., in~\cite{3}.

Note that the time-slot correlation method re-introduces additional aliasing and filtering. It has often been overlooked that by dividing the time axis into finite \textit{regularly spaced} time slots, \textit{one is effectively re-introducing aliasing} into the spectral estimate while also averaging (low-pass filtering) across the introduced slot widths. Special care must be taken when using this method to compute power spectra.\\

\section[Experimental setups] {Experimental setups}
\label{sec:4}

Algorithms based on equations~(\ref{eq:DFT}) and (\ref{eq:directestimatorwkg}) were applied to the burst mode data from two separate experiments: the far axisymmetric turbulent jet and the two-dimensional cylinder wake of~\cite{12}.  The data were also processed using the time-slot approximation of equation~(\ref{eq:ac}), with and without additional anti-aliasing windows. \\

\subsection{The Jet}

Measurements were performed in the streamwise direction 30 jet exit diameters downstream of the jet exit nozzle. This was done at three positions across the jet core: at the center (0\,mm), and at two positions 26 and 52\,mm off the center axis. The exit velocity chosen was U$\,= \,30$\,m/s and the data rate at the center axis position was on average about 134\,Hz. For the time slot correlation spectra and one sample-and-hold spectrum an additional dataset with an average datarate of about 6\,400\,Hz was used for practical reasons, convenience and clarity. Using a dataset with a 134\,Hz average datarate for the time slot based methods would require quite impractical amounts of time to acquire enough data and compute the spectra. Note that the higher datarate will not impact the resulting spectra adversely, but if anything rather to their advantage.

The LDA system was placed to the side of the jet to minimize obstructions of the flow. The optical head was placed on a 3-axis traversing system and connected to the laser through fiber optics. The LDA consisted of Dantec two-component FiberFlow optics mounted on a two-axis traverse, two BSA enhanced processors, and BSA Flow Software version 2.12. A 5\,W Coherent Innova 90 Argon-Ion laser was used, running at 2\,W. Only the 514.5\,nm wavelength was used, measuring the velocity component in the main flow direction. The system was operating in back scatter mode.

The jet was a cubic box of dimensions 58$\times$58.5$\times$59\,cm$^3$ fitted with an axisymmetric plexiglas nozzle, tooled into a 5th order polynomial contraction from an interior diameter of 6\,cm to an exit diameter of d$\, = \,$1\,cm. The interior of the box was stacked with foam baffles in order to damp out disturbances from the fan that supplied the generator with pressurized air. The air intake was located inside the jet enclosure. For further details on the generator box, see Gamard~\cite{Jung,Gamard}. The flow generating box rested on an aluminum frame. The exit velocity was monitored via a pressure tap in the nozzle positioned upstream of the contraction and connected to a digital manometer by a silicon tube. The ambient pressure was monitored by an independent barometer. The enclosure utilized in the experiment was a large tent of dimension 2.5$\times$3$\times$10\,m$^3$. The jet was the same as that used by Ewing \textit{et al.}~\cite{9} and the enclosure was similar as well. Under these conditions, the jet flow generated in the facility should be expected to correspond to a free jet up until x/D$\,=\,70$, which was sufficient for our purpose.

\subsection{The Cylinder Wake}

In this experiment HWA data and LDA data including residence times were acquired under similar conditions in the wake of a circular cylinder in a wind tunnel at DTU\footnote{These were kindly provided by Dr. Holger Nobach at Max-Planck Institut f\"{u}r Dynamik und Selbstorganisation. The HWA data set was downloaded from http://ldvproc.nambis.de/data/dtudata.html, while the LDA data were obtained directly from Holger Nobach.}. The experimental setup for both is approximately the same as described by~\cite{12}. However, some important details are corrected and highlighted in the following. The circular cylinder had a diameter of 6\,mm and was mounted in a square test section of width 300\,mm in a closed loop wind tunnel. The material and mounting (firm or flexible) of the cylinder is unknown; however, it is known that it is some kind of metal and that the rod is solid (as opposed to hollow). The HWA and LDA time series were acquired at the same position 26\,mm downstream of the cylinder. The HWA measurements were acquired at a Reynolds number of 7\,400 ($U = 18.5$\,m/s) based on the cylinder diameter with a sampling frequency of $f_s = 100$\,kHz. The Reynolds number for the LDA data is higher, 8\,600 ($U = 21.5$\,m/s) with an average data rate that is close to that of the HWA measurements, $\nu = 94$\,kHz. Thus the ratio of the mean velocities of the HWA and LDA data sets is $U_{\mathrm{LDA}}/U_{\mathrm{CTA}} = 1.16$, which is important to bear in mind when making comparisons. The spectra computed from these two data sets are compared despite the fact that they are acquired under slightly different conditions.

\section{Results}
\label{sec:5results}

\subsection{Direct estimator}

\begin{figure*}[!h]
\begin{minipage}{0.5\linewidth}
  \center{\includegraphics[width=0.95\textwidth]{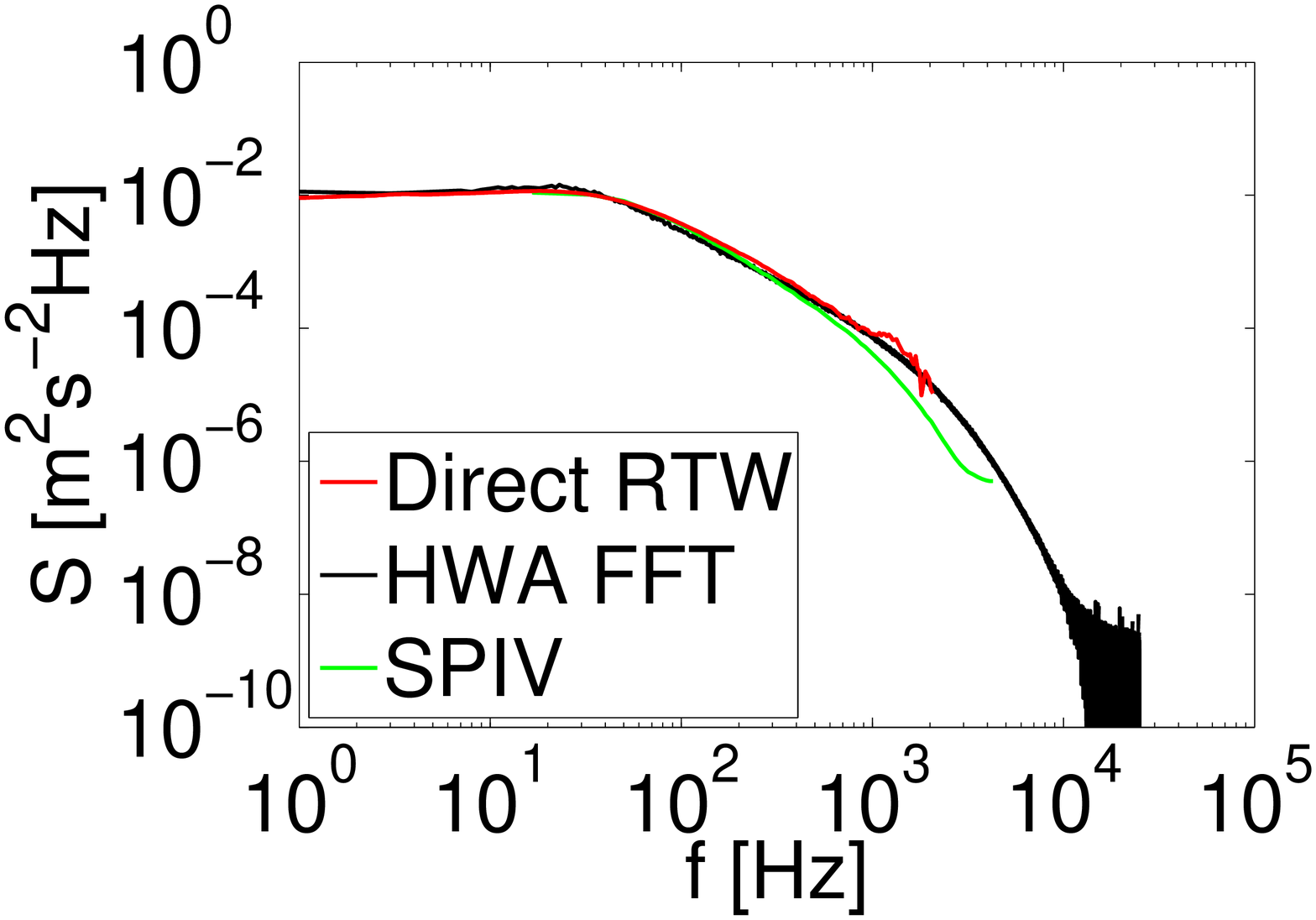}}
\caption{Comparison of burst-mode LDA, HWA and PIV spectra, axisymmetric jet.}
\label{fig:23}
\end{minipage}\hspace{0.25cm}
\begin{minipage}{0.5\linewidth}
  \center{\includegraphics[width=0.95\textwidth]{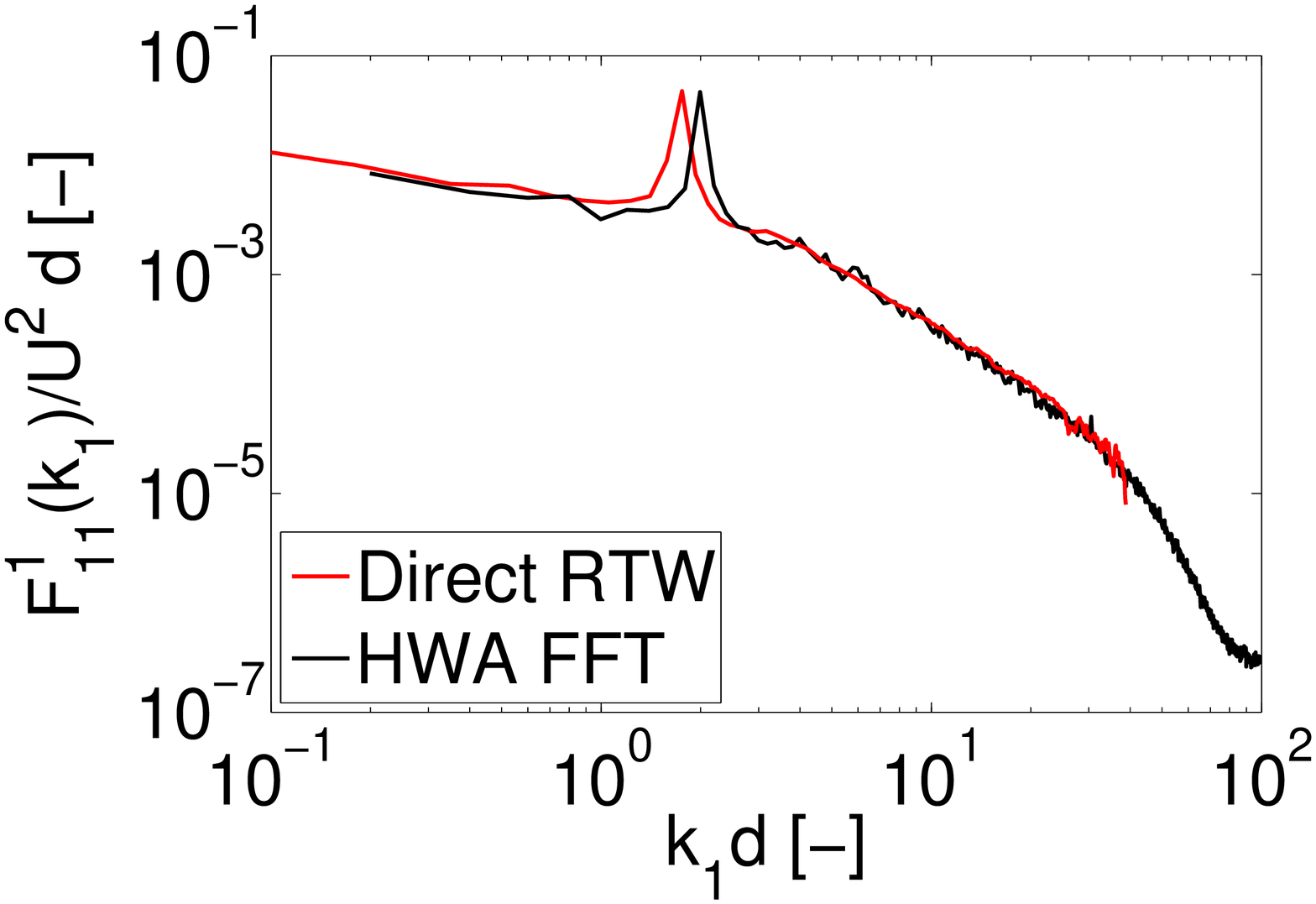}}
\caption{Comparison of burst-mode LDA and HWA spectra, cylinder wake.}
\label{fig:45}
\end{minipage}
\end{figure*}

Figure~\ref{fig:23} shows the direct residence time weighted (RTW) LDA spectrum for the axisymmetric jet along with a spectrum from corresponding HWA data. Also shown is the corresponding spatial spectrum from the Stereoscopic Particle Image Velocimetry (SPIV) data of W\"{a}nstr\"{o}m~\cite{Wanstromthesis,Wanstrometal}, converted from wavenumber to frequency space.\footnote{These data were acquired in the same facility using SPIV streamwise planes which spanned from 30 to 100 diameters downstream. The data were transformed into logarithmic coordinates as described in \cite{9,Wanstromthesis,Wanstrometal}.  The spatial resolution was low because of the large spatial extent of the field.}  It is clear that the SPIV spectrum has an earlier roll-off than the HWA spectrum due to the larger spatial extent of the measuring volume. A constant offset has been removed from the LDA spectrum to ease comparison with the other spectra presented, since the spectral content is otherwise essentially buried in self-noise. The residence time weighted spectrum agrees well with both the HWA and PIV spectra up to a frequency of about 1\,kHz, where the variability and dead-time effects begin to dominate the burst-mode LDA spectrum.

Figure~\ref{fig:45} shows the wavenumber spectra (computed using Taylor's frozen field hypothesis) for the cylinder wake flow. Wavenumber spectra were used since the LDA and HWA measurements were, as mentioned above, acquired at slightly different Reynolds numbers. The reason why the spectral peaks do not line up is that the disturbance is temporal, not convected. This frequency corresponds to the eigenfrequency of the second vibrational mode of the solid cylinder~\cite{4}. The wavenumber spectra are shown to remove the effect of different free stream velocities if the spectra are normalized by the energy by similarity scaling for low wavenumbers (energy variables), i.e., by the factor $u^2L \sim U^2d$~\cite{4}. Again, the direct residence time weighted LDA spectrum is shown to predict the spectrum well, in spite of the fact that the turbulence intensity is relatively high. Since the variability is not as dominating at the higher frequencies for this case (i.e., smaller values of the spectral estimator) due to the increased data rate and more blocks of data, and the dead time effects do not begin to dominate until much higher frequencies, it is possible to resolve more decades of the spectrum (see equation~\ref{eq:converg}).

\subsection{Time Slot Correlation (TSC) method}
\label{sec:4c}

For the sake of completion, the direct estimator is compared to the one obtained using the slotted time lag technique of Buchhave~\cite{3}. The same block length has been used for the direct and time slot correlation method, respectively. The temporal resolution of each respective autocovariance estimate is optimized in relation to, and naturally cannot exceed, the ultimately limiting temporal resolution of the original raw data itself without obtaining a distorted ACF. In the jet case, for practical purposes such as better statistical convergence and computing time, a data set of data rate $\nu \approx 6\,400$\,Hz was used instead of the original $\nu = 134$\,Hz.

The time slot correlation method is (in principle) equivalent to the direct estimator, but as noted earlier these spectra need anti-aliasing filters (since aliasing has been re-introduced by the time-slots). Figures~\ref{fig:Timeslot_spec_linlinW} and~\ref{fig:Timeslot_spec_linlinJ} show the autocovariance (a) and power spectra (b,c) for the jet and the wake, respectively.  As expected and as noted below, these spectra clearly show the leakage to higher frequencies.


\subsubsection{Cylinder wake - temporal resolution and windowing}

Windowing is an unavoidable effect when working with measured records since they have to be finite in time for practical reasons. The windowing effect behaves very differently when computing spectra using the direct and time-slot methods. To illustrate this, Figure~\ref{fig:Timeslot_spec_linlinW}(a) shows the cylinder wake autocovariance of the time slot correlation method as well as its Hanning (0.5 + 0.5\,$\cos (\pi |\tau| / T)$) and Bartlett ($1-|\tau|/T$) windowed counterparts along with autocovariance obtained from inversely Fourier transforming the direct spectral estimates.

This example shows quite nicely how the window function works. From theory (see, e.g., \cite{1,3}) it is well known that applying the Bartlett window on the time-slot data should look exactly like the autocovariance produced from inverse Fourier transformation of the direct estimator, which in this case is seen to have a slightly different period than the other signals. This discrepancy is due to too low a resolution in the spectrum to properly resolve the peak, and is easily remedied by computing the direct estimator with a higher resolution before applying the inverse Fourier transform to it. This new curve falls nicely on top of the Bartlett windowed time-slot correlation curve. The resolution problem introduced by creating the slots, of course, goes both ways but is bypassed when computing spectra directly, where one can decide the frequency resolution completely arbitrarily. Note also that the long time correlation periodicity of the autocovariance in Figure~\ref{fig:Timeslot_spec_linlinW}(a) provides even further support that the periodic disturbance is temporal, and not convected.

In the frequency domain, the spectral leakage of energy from lower to higher frequencies due to windowing is clearly displayed when comparing the direct and time-slot correlated estimates. This is expected since for the truncated autocovariance function the Fourier transform to obtain the power spectrum yields:
\begin{eqnarray}
S_T(f) & = & \int_{-T/2}^{T/2}e^{-i2\pi f\tau}B_u(\tau)\,d\tau \nonumber \\
& = & \int_{-\infty}^{\infty}e^{-i2\pi f\tau}B_u(\tau)W_{\textrm{top-hat}}(\tau)\,d\tau \nonumber \\
& = & S_u(f)\otimes F.T.\{W_{\textrm{top-hat}}(\tau) \}
\end{eqnarray}
where the Fourier transform of the top hat window
\begin{equation}
F.T.\{W_{\textrm{top-hat}}(\tau)\} = \frac{\sin(\pi fT)}{\pi fT}
\end{equation}
is convolved with the spectrum and thereby smears out the energy across frequency from higher to lower energy in the process commonly referred to as `spectral leakage'. This window can assume negative values and falls off as $f^{-1}$, which is quite evident in Figure~\ref{fig:Timeslot_spec_linlinW}(b). The window effect, introduced when first computing the autocovariance to obtain the spectrum, is obviously important to account for and can be remedied only by increasing the maximum record length in the autocovariance. Note that the statistical convergence gets worse for longer time lags, where the estimate is based on less and less samples.

By comparison, for the direct method~\cite{1}:
\begin{equation}
S_T(f) = \int_{-T}^{T}e^{-i2\pi f\tau}B_u(\tau)\left [ 1-\frac{|\tau|}{T} \right ]\,d\tau
\end{equation}
where the Fourier transform of the time window is given by
\begin{equation}
F.T.\{W_{\textrm{Bartlett}}(\tau)\} = \int_{-T}^{T}e^{-i2\pi f\tau} \left [ 1-\frac{|\tau|}{T} \right ]\, d\tau = T\left [ \frac{sin(\pi fT)}{\pi fT} \right ]^2
\end{equation}
This is the Bartlett window, which has two advantages over the top-hat window; It is always positive or zero and it falls off as $f^{-2}$, and hence produces less spectral leakage (though one needs even sharper windows at the very highest frequencies for turbulent spectra, which roll off even faster).

If the autocovariance converges rapidly to zero, one may consider using zero padding in the time slot approach to decrease the windowing effect on the resulting spectrum. However, any discontinuities will cause ringing of the signal and the approach may not be realistic if the number of required cross products becomes impractically large and the computing time becomes very long, in particular to avoid the $f^{-1}$ roll-off in the spectrum caused by the limited record length windowing. It can further be shown that the variability of the autocovariance is inversely proportional to the value of the autocovariance, so when it approaches zero the variability goes to infinity (see, e.g., \cite{1}). The direct method is considerably faster to compute and the record length is not as critical as long as the record is much longer than the integral time scale. On the other hand, it is inefficient to use lags much longer than those relevant for the scales in the flow. It is therefore a common approach to divide the record into blocks that still include the relevant flow scales and average the resulting spectra.

\subsubsection{Jet - temporal resolution and aliasing}

\begin{figure*}
\begin{minipage}{0.50\linewidth}
  \center{\includegraphics[width=1.0\textwidth]{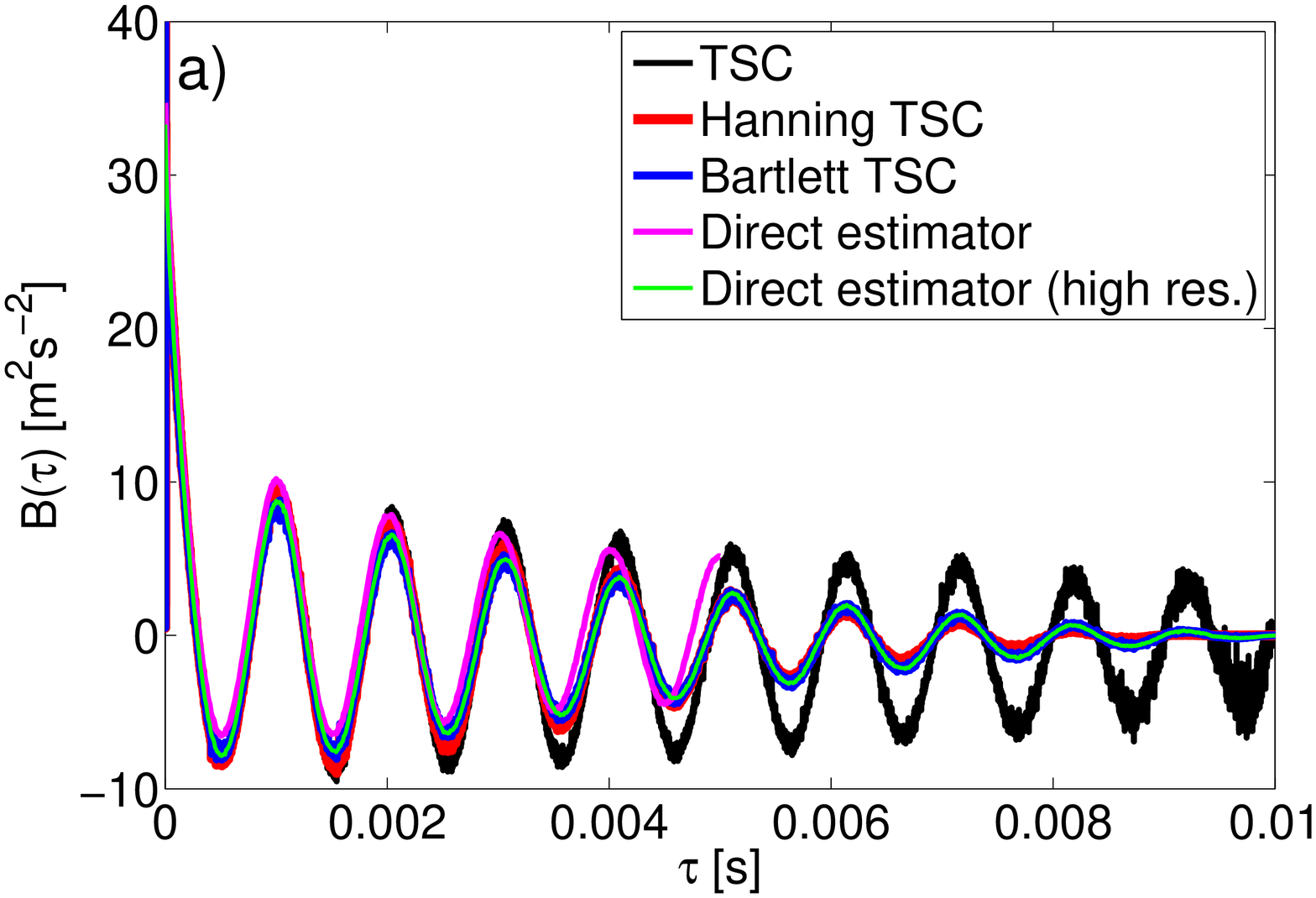}}
\label{fig:Timeslot_AC}
\end{minipage}\hspace{0.5cm}
\begin{minipage}{0.50\linewidth}
  \center{\includegraphics[width=1.0\textwidth]{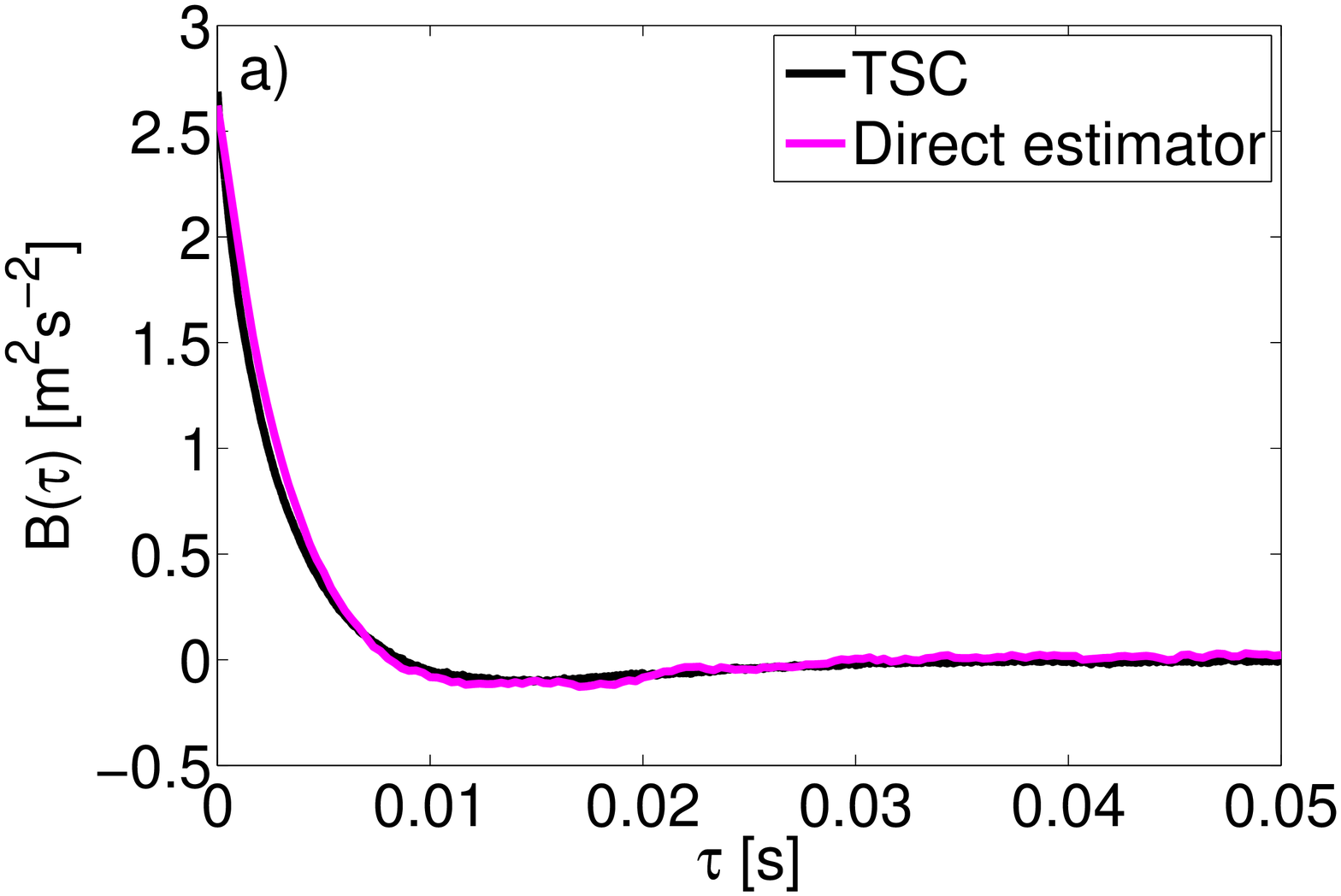}}
\label{fig:Timeslot_AC}
\end{minipage}\\
\begin{minipage}{0.50\linewidth}
  \center{\includegraphics[width=1.0\textwidth]{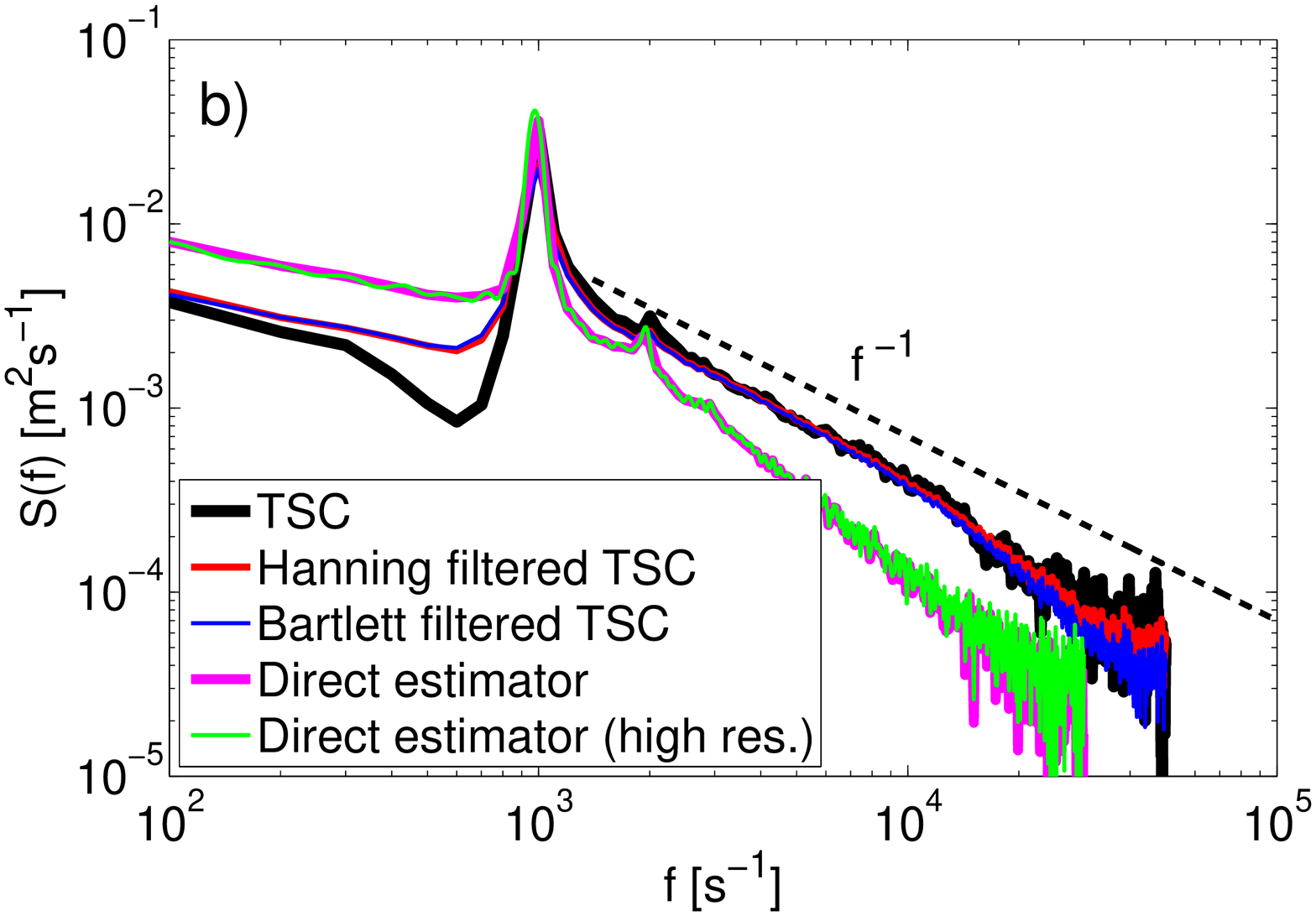}}
\label{fig:Timeslot_spec_loglog}
\end{minipage}\hspace{0.5cm}
\begin{minipage}{0.50\linewidth}
  \center{\includegraphics[width=1.0\textwidth]{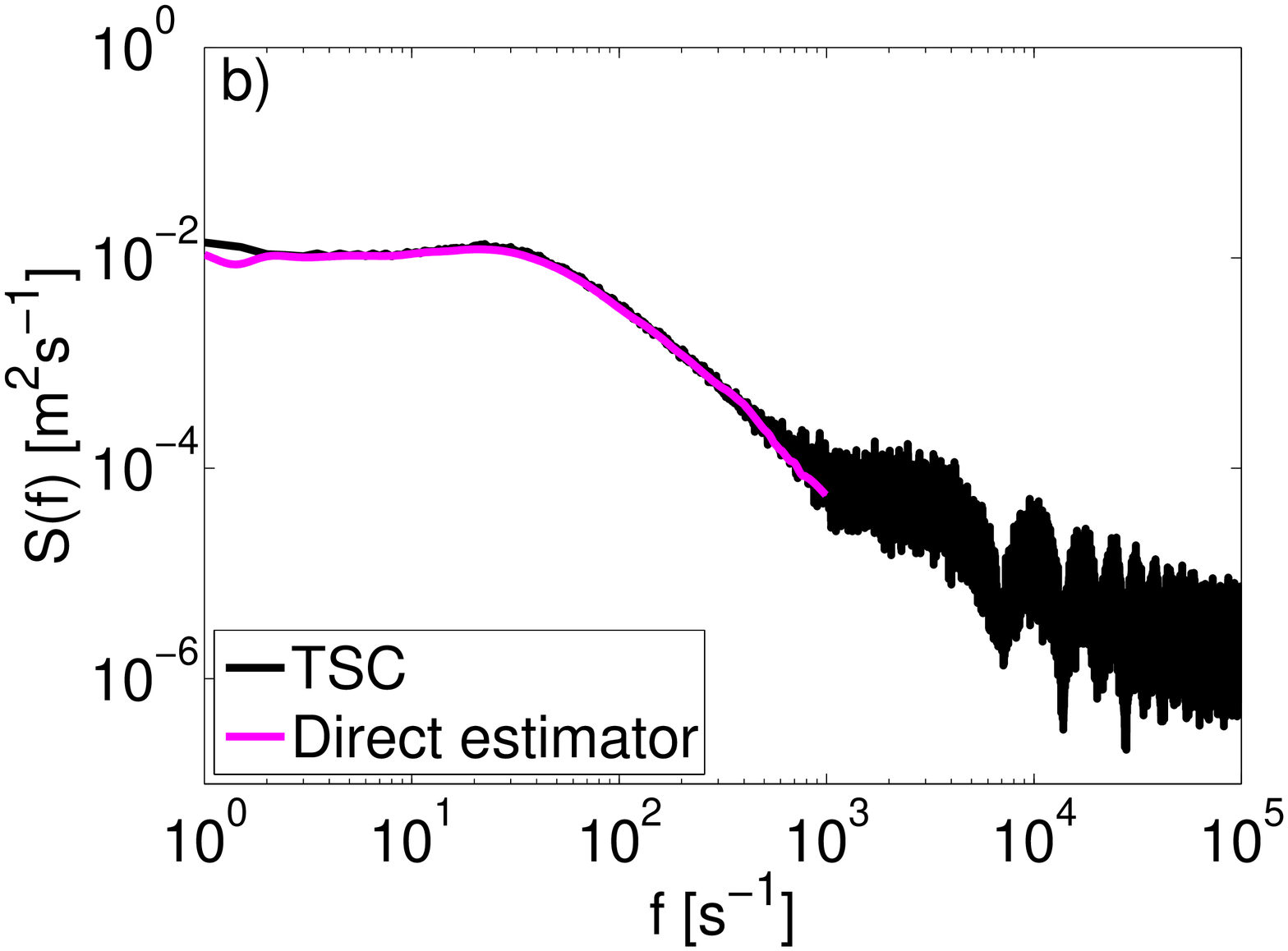}}
\label{fig:Timeslot_spec_loglog}
\end{minipage}\\
\begin{minipage}[t][10cm][t]{0.50\linewidth}
  \center{\includegraphics[width=1.0\textwidth]{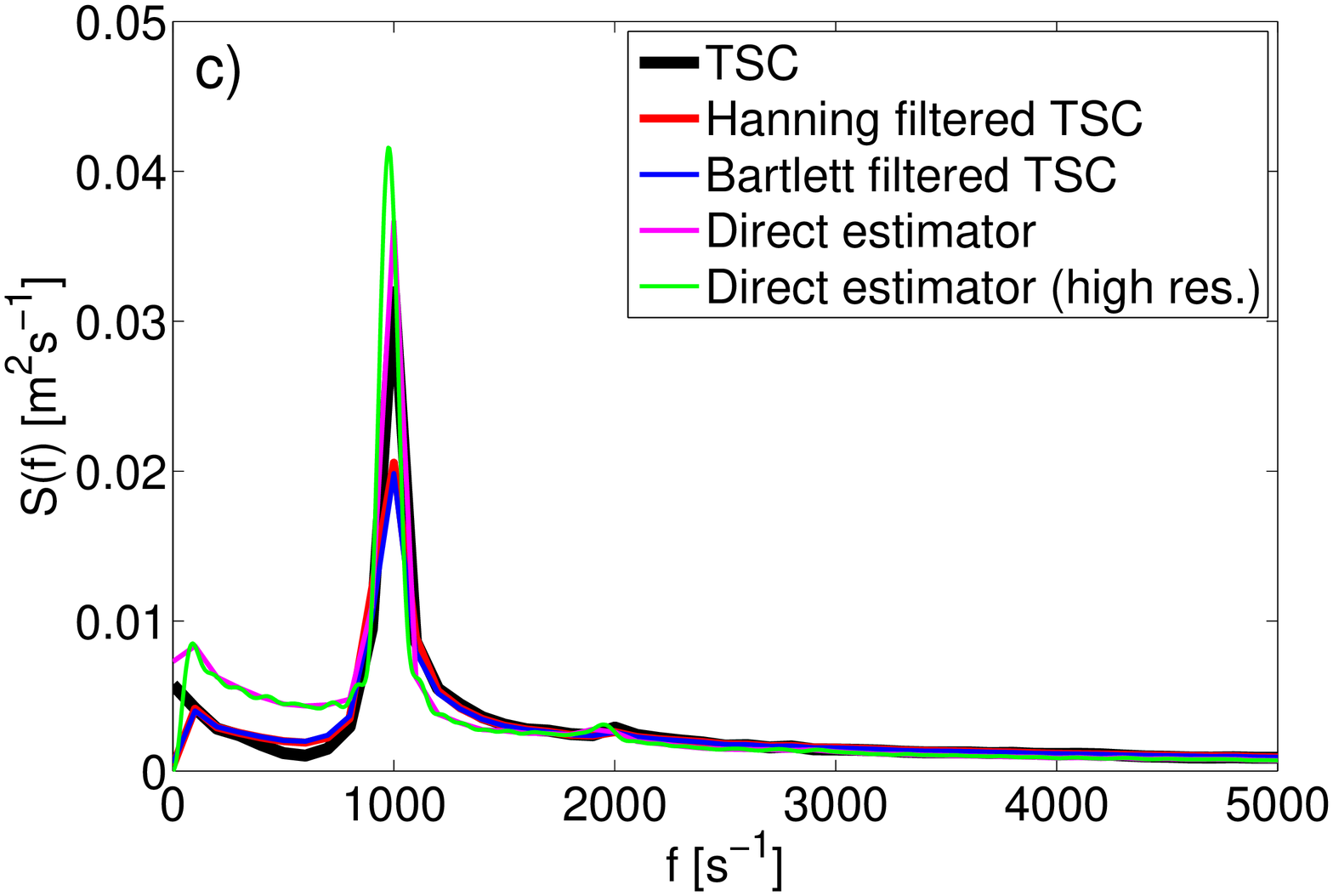}}
\caption{Cylinder wake: Autocovariance (top) and spectra (middle and bottom) from the time slot method (black), the same with a Hanning window (red) and a Bartlett window (blue) imposed. For the sake of comparison, we also display the autocovariances computed by the inverse Fourier transform of the direct estimator for a lower frequency resolution of $\Delta f =$ 100\,Hz (magenta) and the same with a higher frequency resolution of $\Delta f =$ 5\,Hz (green).}\label{fig:Timeslot_spec_linlinW}
\end{minipage}\hspace{0.5cm}
\begin{minipage}[t][10cm][t]{0.50\linewidth}
  \center{\includegraphics[width=1.0\textwidth]{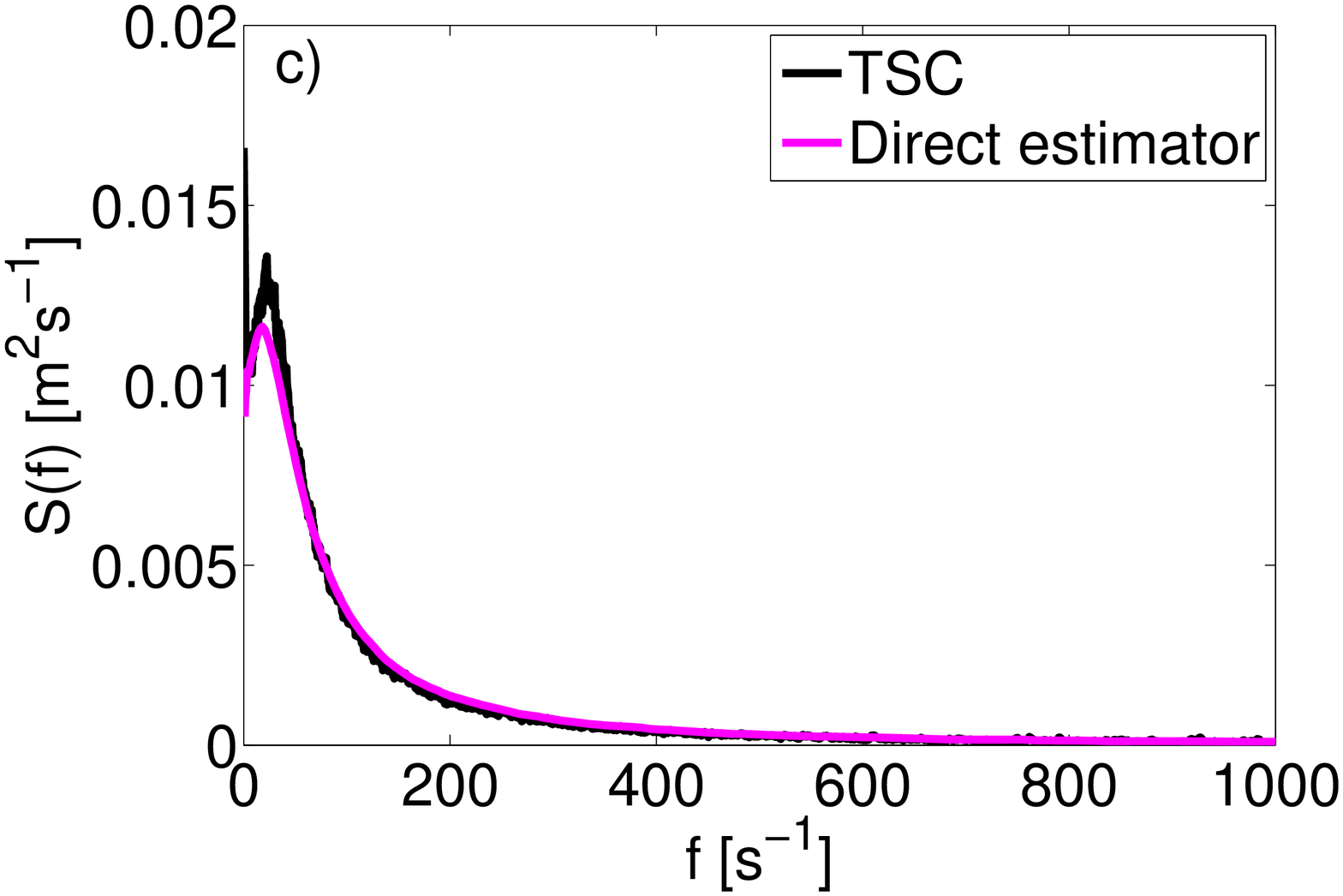}}
\caption{Axisymmetric jet: Autocovariance (top) and spectra (middle and bottom) from the time slot method (black) and the direct estimator (magenta).} \label{fig:Timeslot_spec_linlinJ}
\end{minipage}\hspace{0.5cm}
\end{figure*}

Figures~\ref{fig:Timeslot_spec_linlinJ}(a-c) show corresponding results for the turbulent axisymmetric jet. The oscillations at higher frequencies are a direct result of the spectral leakage due to the finite record length window.


\subsection{Other alternative algorithms} \label{sec:5main}

\begin{figure*}
\begin{minipage}{0.5\linewidth}
  \center{\includegraphics[width=0.95\textwidth]{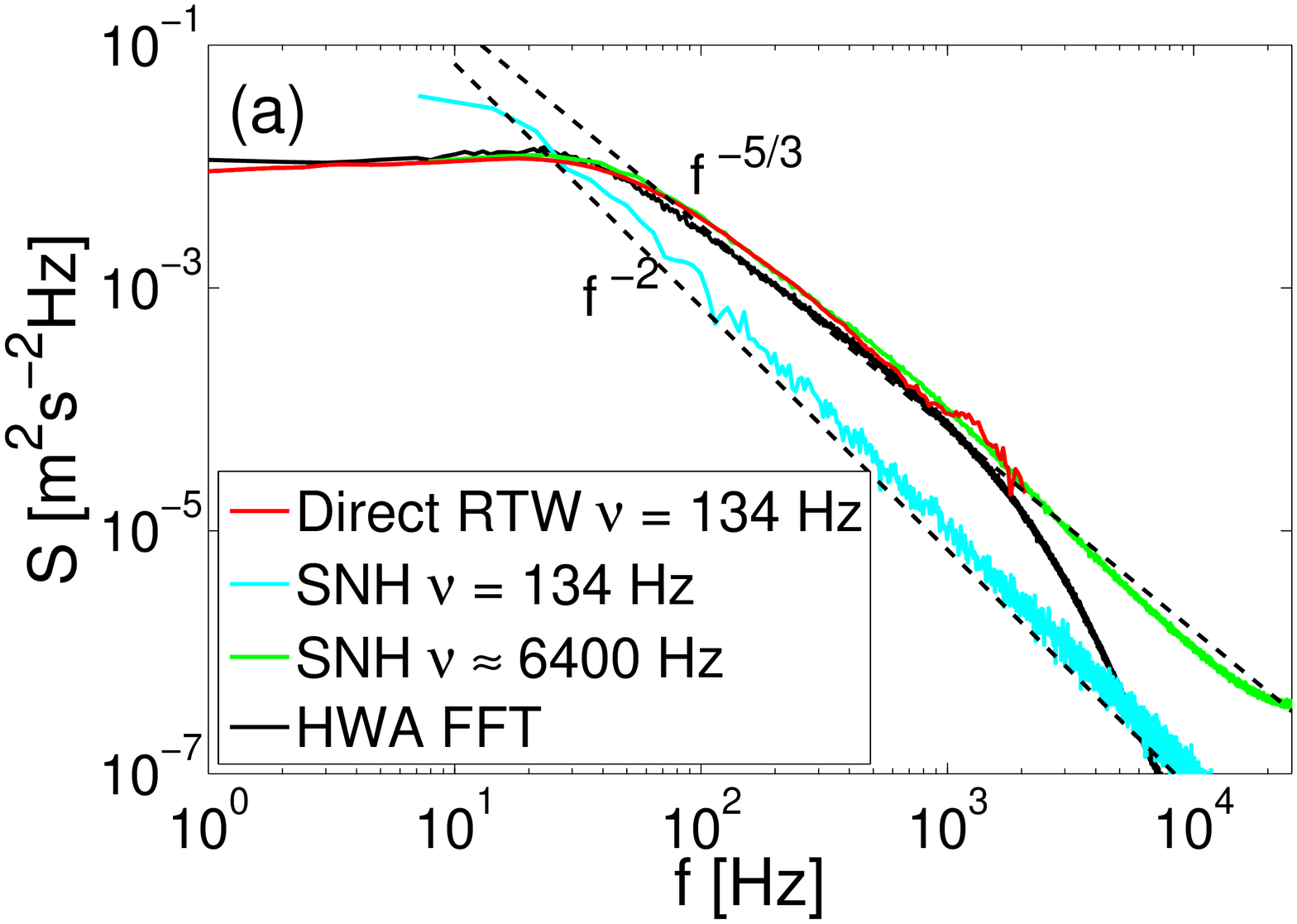}}
\end{minipage}
\begin{minipage}{0.5\linewidth}
  \center{\includegraphics[width=0.95\textwidth]{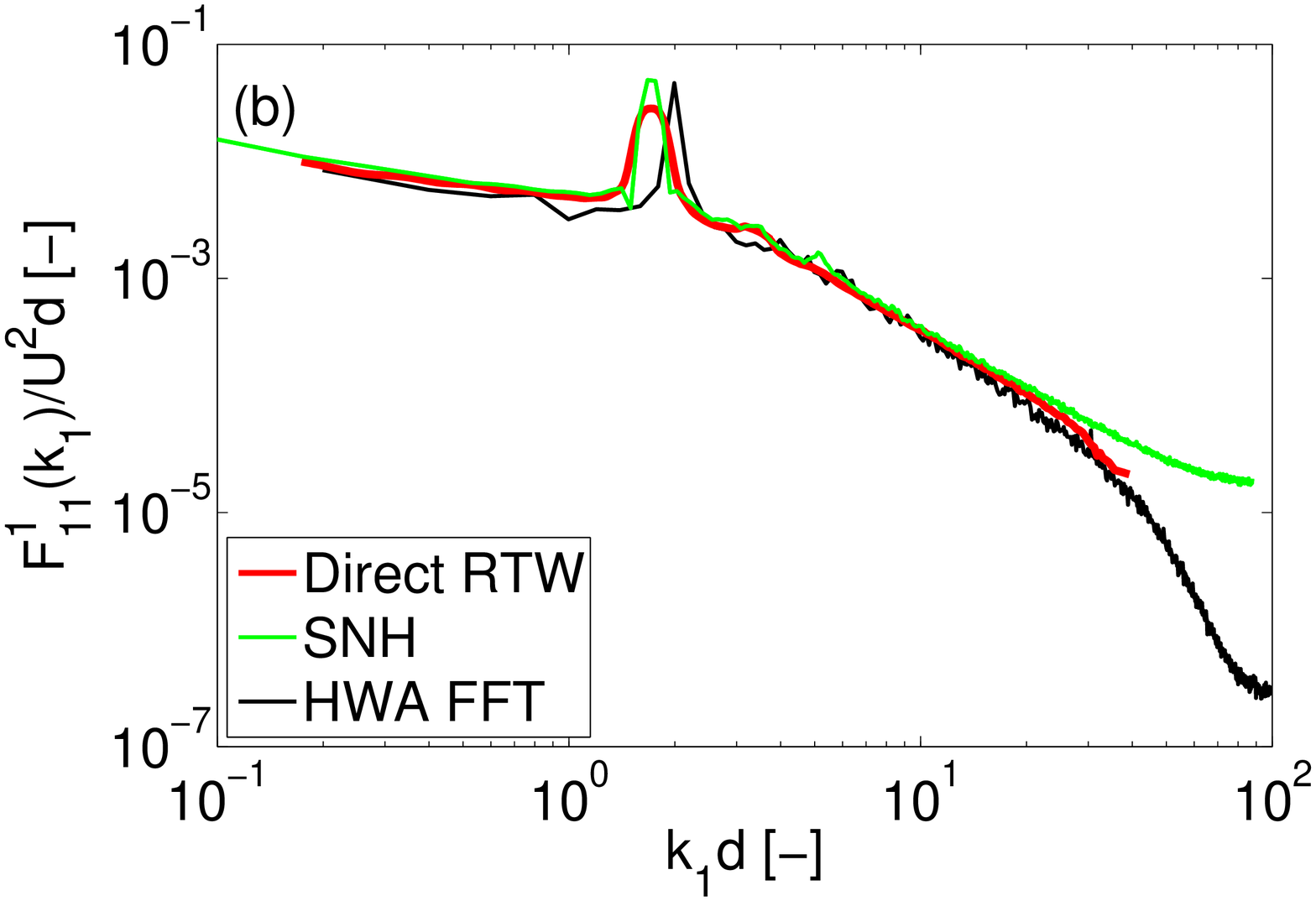}}
\end{minipage}
\caption{Comparison to alternative algorithms for the (a) turbulent axisymmetric jet ($\nu = 134$ and $6\,400$\,Hz with $\nu \lambda_{\tau}=0.22$ and $15.4$, respectively) and (b) cylinder wake data ($\nu =94$\,kHz with $\nu \lambda_{\tau}=94$).}
\label{fig:456}
\end{figure*}

\subsubsection{Interpolation and re-sampling} \label{sec:5}

Also shown for comparison in Figures~\ref{fig:456}(a,b) are
spectra computed from the same data using standard FFT analysis on
re-sampled data obtained from a continuous signal created by
sample-and-hold (SNH) from the same LDA burst-mode data. For the
axisymmetric jet, the original data set with data rate $\nu = 134$\,Hz are supplemented with a dataset of $\nu \approx 6\,400$\,Hz.
The low data rate sample-and-hold spectrum is very different
from both the hot-wire one and the one obtained from the LDA data
using the residence-time burst-mode method described above. By contrast, the high data rate one performs quite well. Since the probe volume is smaller and a high data rate can better resolve smaller flow scales, this method may, under favorable conditions, perform even better than the HWA. Note, however, that simply increasing the data rate is not always a successful strategy since this may compromise the data quality, as further discussed below. Further, as in figure~\ref{fig:456}, the highest frequencies of the SNH spectra may of course be unreliable since the method naturally cannot capture the effects beyond the finite probe volume even though they can still be computed.

For the wake, there is very little difference except at high frequencies and at the very lowest frequencies. This is consistent with the analysis
of Adrian and Yao~\cite{13}, since the data density, $\nu
\lambda_{\tau}$ ($\nu$ is the data rate and $\lambda_{\tau}$ is the temporal Taylor microscale), is a measure of how many samples on average are
taken during one Taylor microscale. This parameter is clearly critical for the performance of interpolation
and re-sampling methods, as was pointed out in~\cite{3,4,13} and is also flagrant in Figures~\ref{fig:456}(a,b). For the
axisymmetric jet, $\nu \lambda_{\tau}=0.22$ and $15.4$ for $\nu = 134$\,Hz and $6\,400$\,Hz, respectively. Therefore the re-sampled
data cannot represent the signal nearly as well in the former case as it does in the latter~\cite{4}. Unless $\nu
\lambda_{\tau} \gg 1$, the spectra will be subject to frequency
dependent noise, falling off as $f^{-2}$~\cite{3,13} for
sample-and-hold, which is evident in Figure~\ref{fig:456}(a). Note
that, if not looking too carefully, it would have been very easy to
have confused the sample-and-hold spectrum with that of turbulence (since -2 is close to -5/3),
when in fact the spectral shape is imposed
upon it by the reconstruction. The spectrum is affected adversely even at the lowest frequencies, where the energy is over-predicted. For the cylinder wake, the average sampling rate is substantially higher with a much higher data
density $\nu \lambda_{\tau}=94$, so a better representation of the
instantaneous signal is possible~\cite{4}.

One can clearly see that, in contrast to the sample-and-hold
spectra, the burst-mode LDA spectra collapse with the HWA spectra \textit{independently} from data density. This is in direct contrast
to the high data density ($\nu \lambda_{\tau} \gg 1$) required for the re-sampling methods to
work. High data density in comparison to the flow scales is typically a very difficult
condition which is rarely satisfied for LDA measurements, and
failure to recognize the problem can seriously compromise the
quality of the acquired data. Moreover, attempts to satisfy it
experimentally risks violate the requirement of at most one
particle at a time in the scattering volume. This would make it
behave like a tracker, producing ambiguity noise~\cite{5}.

\subsubsection{Alternative weighting schemes} \label{sec:AWEs}
As was also established in section~\ref{sec:1}, residence time weighting has previously been shown theoretically~\cite{3} and experimentally~\cite{4} to be the in-principle correct way of computing power spectra (and all other flow statistics) from burst-mode LDA data. Note that the only assumptions for the residence time weighting method is that the particles are homogeneously distributed at one initial instant and that there is at most one particle in the measuring volume at a time. However, the validity of this method has been and still is disputed, and alternative methods have been proposed. Partly on theoretical grounds, partly with reference to practical problems with the measurement of the residence time. However, residence time weighting is still considered the only correct way to process the statistics, requiring that the flow is spatially homogeneously seeded~\cite{2,5,6a}. This obviously requires that the burst processor provides correct and reliable values of the residence times. Some strategies to check this are further discussed in section~\ref{sec:7}.

\begin{figure*}
\begin{minipage}{0.5\linewidth}
  \center{\includegraphics[width=1.0\textwidth]{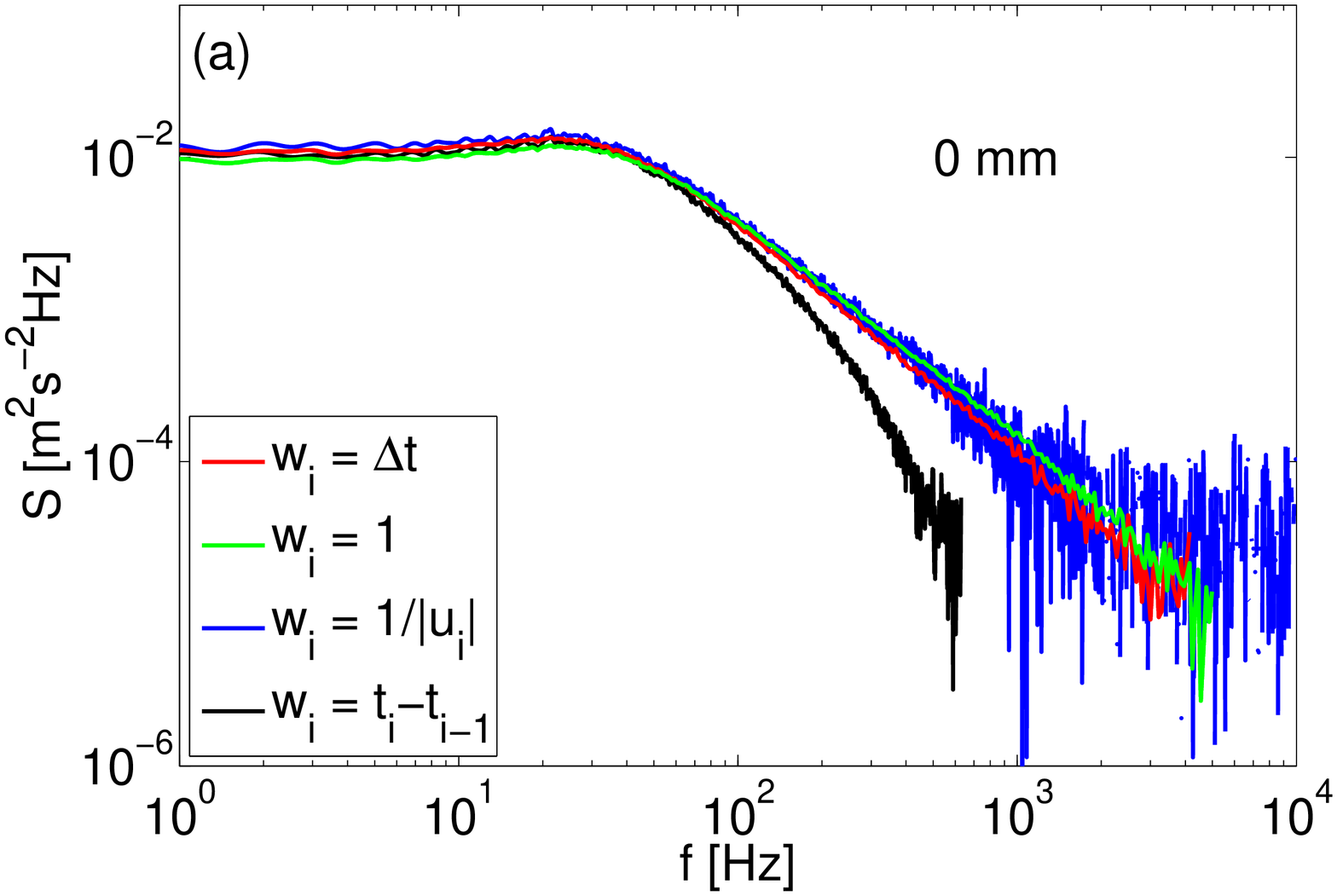}}
\end{minipage}\hspace{0.5cm}
\begin{minipage}{0.50\linewidth}
  \center{\includegraphics[width=1.0\textwidth]{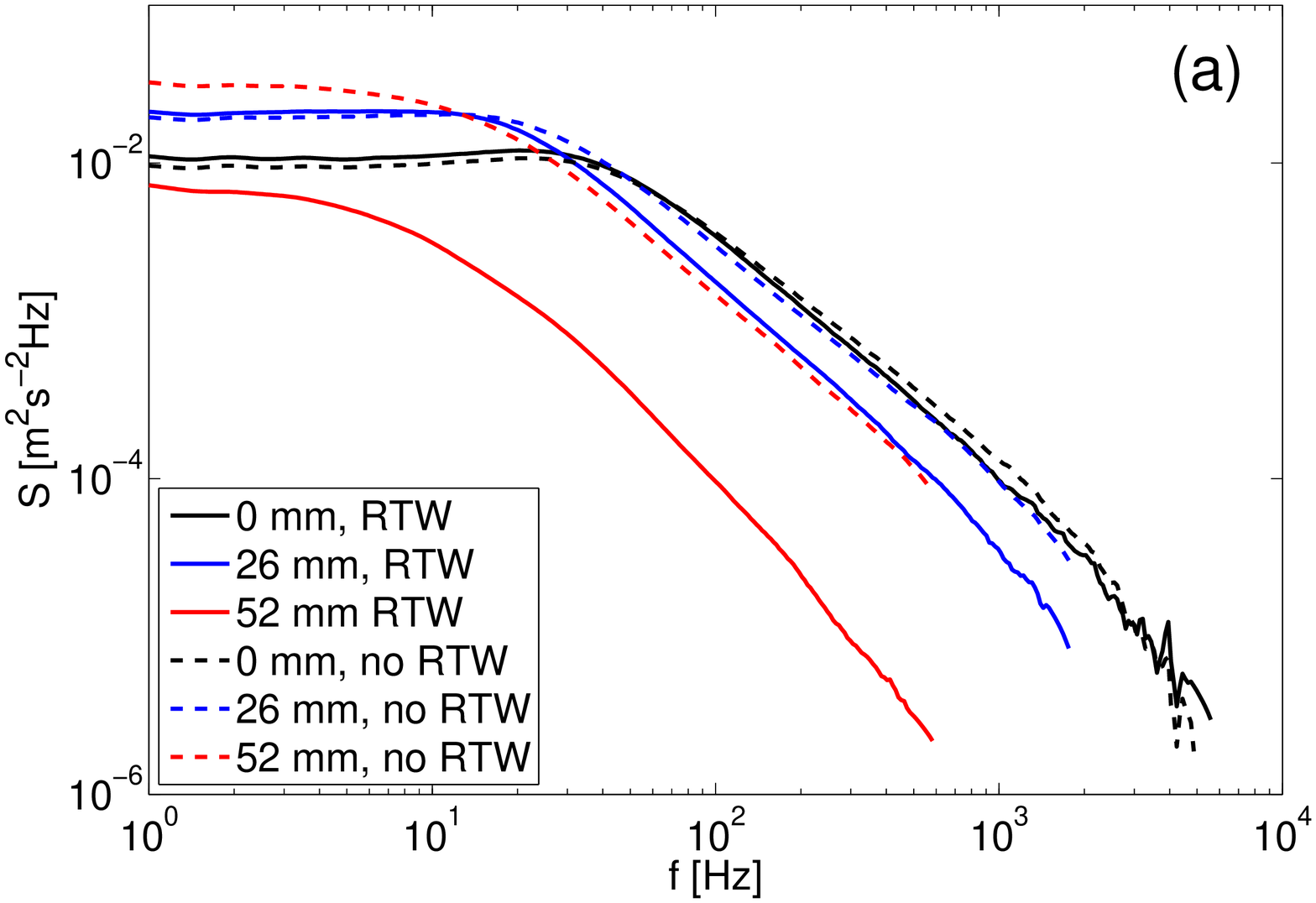}}
\label{fig:offaxisspectra}
\end{minipage}\\
\begin{minipage}{0.5\linewidth}
  \center{\includegraphics[width=1.0\textwidth]{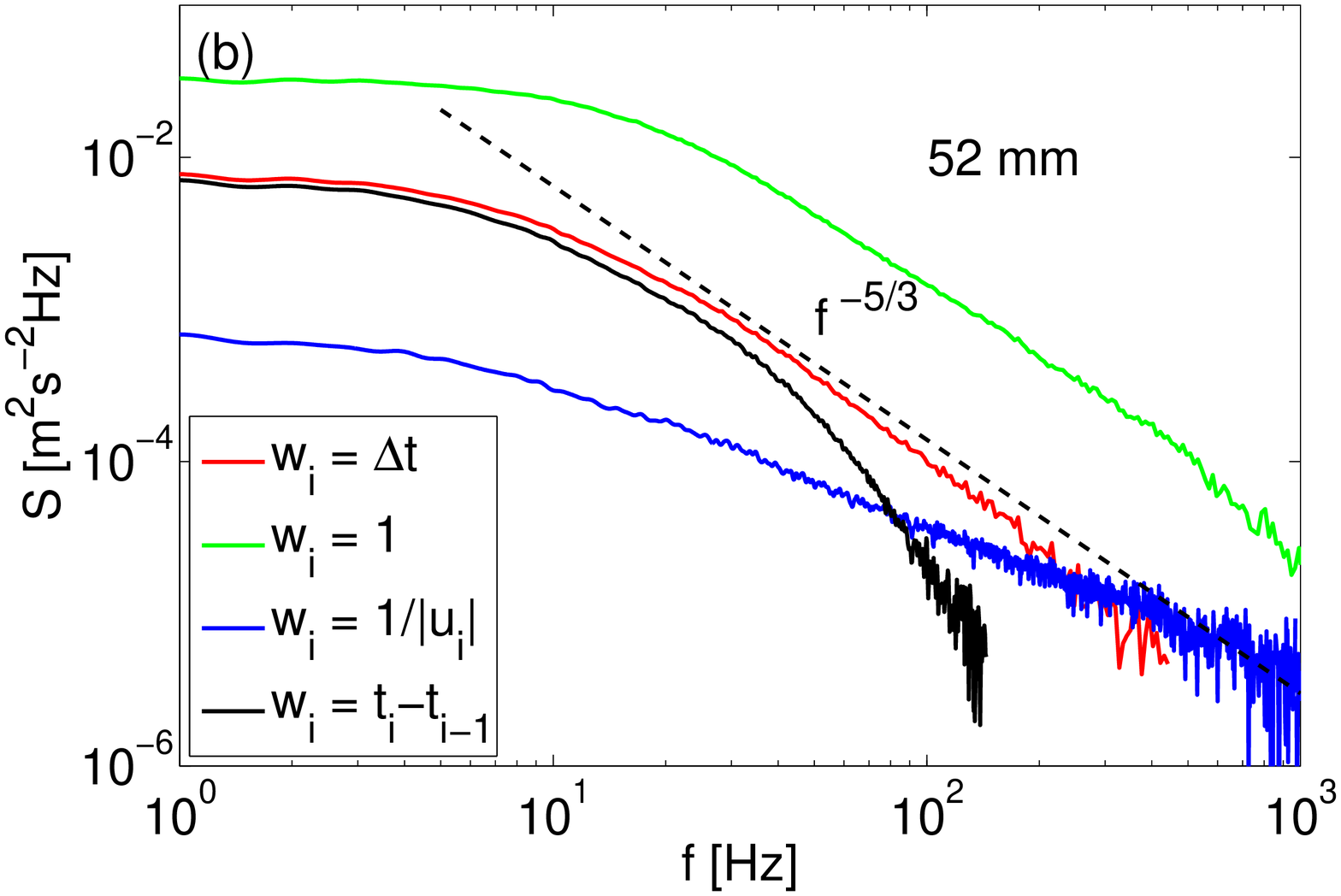}}
\end{minipage}\hspace{0.5cm}
\begin{minipage}{0.50\linewidth}
  \center{\includegraphics[width=1.0\textwidth]{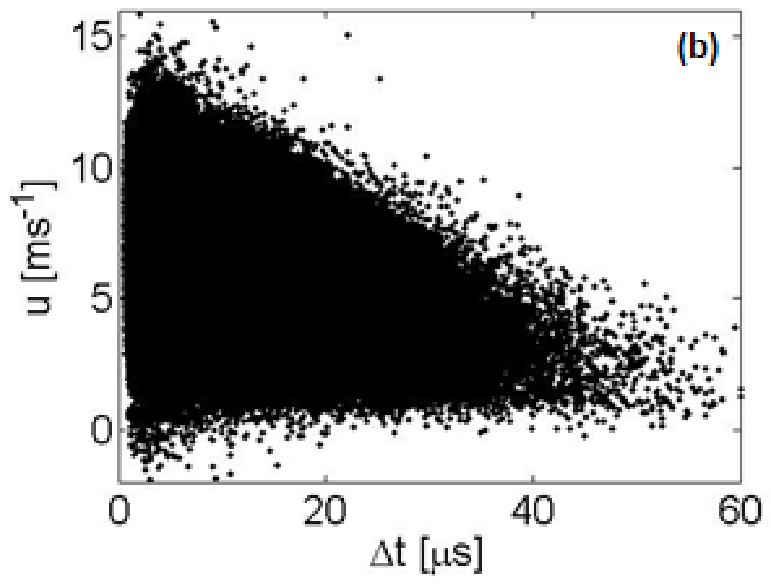}}
\label{fig:Timeslot_spec_loglog}
\end{minipage}\\
\begin{minipage}[t][10cm][t]{0.50\linewidth}
  \center{\includegraphics[width=1.0\textwidth]{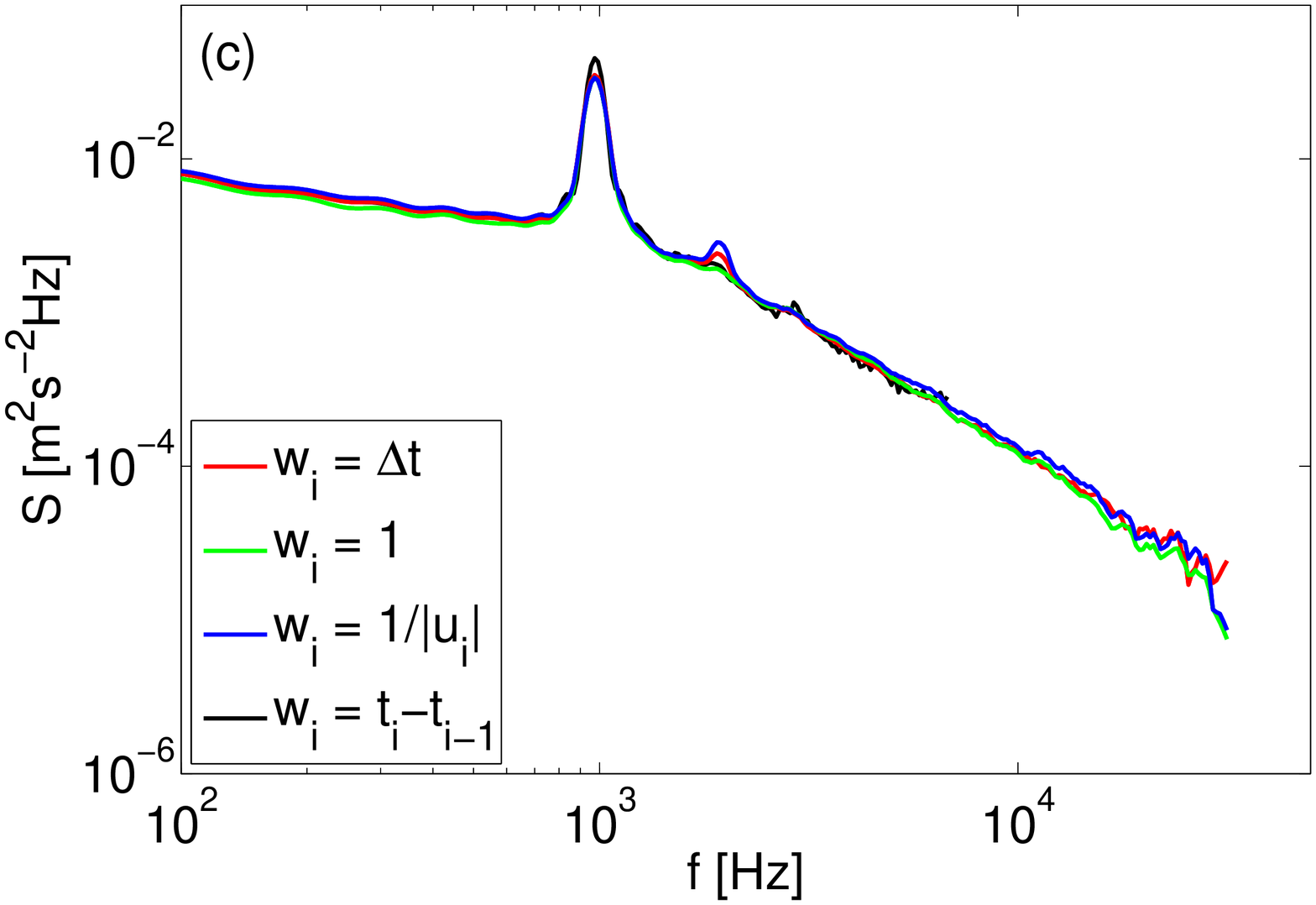}}
\caption{Comparison to alternative weightings for the (a) turbulent axisymmetric jet at the center axis and (b) at 52\,mm off the center axis and (c) the cylinder wake data.} \label{fig:weightings}
\end{minipage}\hspace{0.5cm}
\begin{minipage}[t][10cm][t]{0.50\linewidth}
  \center{\includegraphics[width=1.0\textwidth]{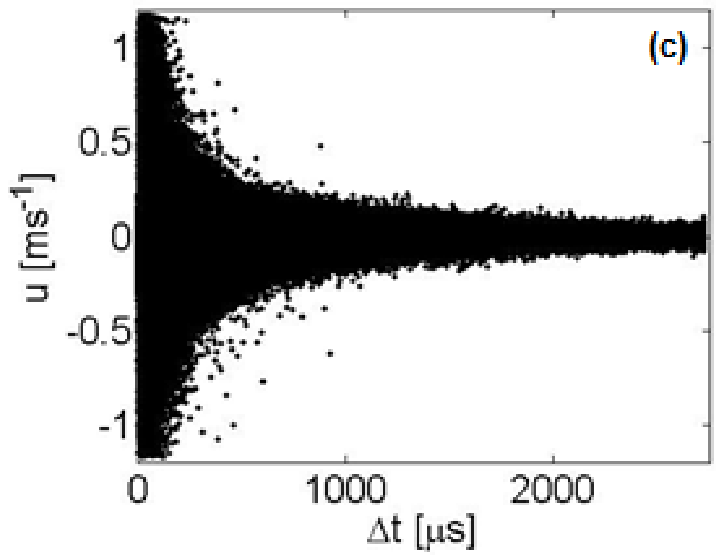}}
\caption{Axisymmetric jet: (a) Off-axis spectra demonstrating the effect of omitting the residence time weighting. Velocity/residence time scatter plots for (b) y\,$=$\,0\,mm and (b) y\,$=$\,54\,mm} \label{fig:offaxis}
\end{minipage}
\end{figure*}

Other traditional weighting schemes include using the inverse absolute velocity (1/$|u_i|$), interarrival times ($t_i-t_{i-1}$) and the free running processor where the weighting factor is simply set to unity. To show the impact of these alternative weighting schemes, the corresponding spectra have been computed and compared to the residence time weighted one at the jet center axis, 52\,mm off the jet center axis and in the cylinder wake, see Figure~\ref{fig:weightings}. The off-axis jet measurements pose more stringent requirements on the weighting since they provide higher turbulence intensity in a shear flow where the mean velocity approaches zero. Though often encountered in practical flows this type of case seems to be seldomly included in similar tests, perhaps due to the larger obtained discrepancies, but is included here since the authors consider it a more reliable test of the weighting schemes.

As can be seen from the Figures~\ref{fig:weightings}(a-c), for the jet centerline and the cylinder wake most weighting schemes appear to agree fairly well, though logarithmic plots can often be misleading in particular for the slow high frequency roll-off in turbulence spectra. Hence, as a check of the validity of the spectra, one should always compare the integral under the spectrum to the variance obtained directly from the data as stated in equation~(\ref{eq:energy}) since the integral of the power spectrum should equal to the power of the signal by definition. The comparison is presented in Table~\ref{tab:weigtings} (using only half the variance since the integral is taken over the half-line spectrum), which reveals that the jet centerline spectra do not agree as well as figure~\ref{fig:weightings}(a) implies. In fact, the inverse velocity weighting does not flatten out at higher frequencies, leaving a significantly over-predicted energy under the integral of the spectrum. As expected, the differences become even more obvious as one moves away from the jet center axis, see figure~\ref{fig:weightings}(b). The residence time weighting still produces correct values, while it is obvious both from the figure as well as Table~\ref{tab:weigtings} that all the remaining weighting schemes fail miserably.

The cylinder wake has a very high data density and low turbulence intensity, which reduces the impact of the high velocity bias and most weighting schemes.  Therefore, not surprisingly, all perform satisfactorily for this particular set of data.

As indicated earlier, it is obvious that the true test of LDA signal processing is in high turbulence flows with low mean velocities (and therefore low data density) and high shear. Common examples, also highly relevant for engineering flows, are \textit{off} the center axis of the jet and near a wall in a boundary layer. It was therefore considered relevant to study the off-axis jet more in detail in the following section. 

\begin{table}
\caption{Comparison of half-line integrals of spectra of various weighting schemes to half the variance obtained directly from the data.}
\label{tab:weigtings}       
\begin{tabular}{lccccc}
\hline\noalign{\smallskip}
Measurement & $\overline{u^2}/2$ from data  & $\Delta t$    & 1     & $\frac{1}{|u_i|}$ & $t_i-t_{i-1}$ \\
\noalign{\smallskip}\hline\noalign{\smallskip}
Jet 0\,mm   & 1.46                          & 1.44          & 1.51  & 8500              & 1.02  \\
Jet 52\,mm  & 0.12                          & 0.11          & 0.99  & 130               & 0.08  \\
Wake        & 17.1                          & 15.9          & 15.2  & 16.3              & 14.9  \\
\noalign{\smallskip}\hline
\end{tabular}
\end{table}

\subsubsection{Jet off-axis measurements} \label{sec:5aaa}

Though the turbulence intensities are quite high at the jet centerline (25 \%), the bias from omitting the residence time weighting is still relatively small, see Figure~\ref{fig:offaxis}(a).  In part, and as shown in Appendix C, this can be understood by considering how higher order moments affect the measurements.  However, a much more stringent test would be to measure in the outer part of the jet, where the shear becomes significant and the turbulence intensity increases due to the lower convection velocity. As the convection velocity approaches zero, the residence times increase dramatically in length. The difference can be seen in the scatter plots of the velocity against the residence times in Figures~\ref{fig:offaxis}(b) and (c) for the centerline and $y\,=$\,54\,mm off the center axis, respectively. Figure~\ref{fig:offaxis}(a) shows power spectra for these two positions and an intermediate one at $y\,=$\,26\,mm, showing the increased discrepancy between the residence time weighted (RTW) and non-weighted (no RTW) spectra with increased distance $y$ to the center axis.

The effects are of course also present in the statistical moments of the measured velocities. Tables~\ref{tab:1}-\ref{tab:4} show the first three velocity moments for the three measurements presented in figure~\ref{fig:offaxis}(a) with and without residence time weighting. The errors caused by omitting the residence time weighting are seen to increase as one moves further away from the centerline. A brief simplified analysis on these errors based on Taylor expansion can be found in Appendix C.\\

\begin{table}
\caption{Statistical moments using residence time weighting and without, $y = 0$\,mm.}
\label{tab:1}       
\begin{tabular}{lcccc}
\hline\noalign{\smallskip}
Algorithm & $\overline{u}$ & $\overline{u^2}$ & $\overline{u^3}$ \\
\noalign{\smallskip}\hline\noalign{\smallskip}
RTW         & 6.24 & 2.78 & 0.32 \\
no RTW      & 6.64 & 2.69 & 0.12 \\
\noalign{\smallskip}\hline
\end{tabular}
\end{table}

\begin{table}
\caption{Statistical moments using residence time weighting and without, $y = 26$\,mm.}
\label{tab:3}       
\begin{tabular}{lcccc}
\hline\noalign{\smallskip}
Algorithm & $\overline{u}$ & $\overline{u^2}$ & $\overline{u^3}$ \\
\noalign{\smallskip}\hline\noalign{\smallskip}
RTW         & 3.1 & 2.2 & 1.4 \\
no RTW      & 3.7 & 2.3 & 1.1 \\
\noalign{\smallskip}\hline
\end{tabular}
\end{table}

\begin{table}
\caption{Statistical moments using residence time weighting and without, $y = 52$\,mm.}
\label{tab:4}       
\begin{tabular}{lcccc}
\hline\noalign{\smallskip}
Algorithm & $\overline{u}$ & $\overline{u^2}$ & $\overline{u^3}$ \\
\noalign{\smallskip}\hline\noalign{\smallskip}
RTW         & 0.24 & 0.24 & 0.24 \\
no RTW      & 0.55 & 0.62 & 0.58 \\
\noalign{\smallskip}\hline
\end{tabular}
\end{table}

\section{Practical considerations} \label{sec:7}

Obviously it is crucial that whatever method is used to process the
LDA signal burst, it must produce correct values of all three
necessary parameters: $u_{n}$, $\Delta t_n$, and $t_n$. While
both leading commercial processors produce reliable values of at
least the velocity, there have been a variety of problems reported
for $\Delta t_n$ and $t_n$ (e.g.,~\cite{15,16}). The
former~\cite{15} used older Dantec (DISA) counters, most versions of
which (if unmodified) produced random numbers for the residence
times due to overflow of the register used to count the pulses in a
burst. To the best of our ability to determine, the current Dantec
burst processors used herein function correctly. Shiri~\cite{17}
using the same hardware, however, encountered problems with the
particle arrival times, $t_n$, when using very large data sets
($>10\,000$ realizations) with multiple channels -- the data
reported from one channel appeared to skip several realizations,
confusing the arrival times and which pairs of data belonged
together. These were apparently due to buffer overflow, and easy to
spot by comparing the arrival and residence times for both channels
(which should be nearly the same). The cause of the problems
encountered by~\cite{16} using TSI correlation processors were never
really sorted out (to the best of our knowledge), but appeared to be
due to an incorrect reading of the residence time data registers,
and therefore an incorrect reassembling of the binary data.
George~\cite{6} (see also~\cite{4,18}) discuss a number of useful
diagnostics to make sure all these parameters have been correctly
reported by the hardware. One of the most useful of these is to
simply examine scatter plots of $u_{n}$ versus $\Delta t_n$ (see, e.g., Figure~\ref{fig:offaxis}b and c),
which should have a characteristic banana shape for positive convection velocities and moderate turbulence intensity (since one expects short $\Delta t$'s for high velocities and vice versa). Large numbers of
realizations with very small residences times suggest signal
quality/processing is deficient. An alternative strategy (in fact
ultimately used by~\cite{16} to overcome problems with their TSI
processors) is to produce residence time data by simply recycling
the burst processor for a fixed number of fringe-crossings (say 8 or
16) within the same burst as many times as possible, outputting each
sub-burst immediately while recycling the processor to measure for
another 8 or 16 fringe-crossings. The key is to use enough frequency
shift to ensure that both the long and short residence times are
well-represented in the data set. For such measurements, simple
arithmetic averaging is equivalent to the residence-time weighted
results, since the weighting is accounted for by the number of times
a given burst is counted. Fourier analysis can be performed in the
same manner, or by regrouping the data to reconstruct the actual
burst. The latter reduces considerably the amount of data to be
processed. Although functional, this approach is at best a crude fix
to a hardware problem that should be correctable (at least by the
manufacturer).

Due to the inherent random sampling, Fast Fourier Transform
algorithms cannot be used to compute the direct Fourier transform of
the burst mode velocity signal. Hence, the spectrum has to be
computed by the direct Fourier transform (DFT). This can be done by
squaring the result of a single summation for selected frequencies,
or by addressing the double summation directly. The latter approach is more computationally expensive  since it
requires a double sum to be performed over the entire block of data
for each frequency. Matlab performs poorly in this situation, since
it needs to re-allocate memory for every loop iteration. FORTRAN
does not have this problem, since memory allocation for each element
is done before entering the loops. But even when parallelizing the
code, the double summation computational time can be hours or days
for the data sets presented here, depending on record lengths and
number of blocks used.

An alternative approach is to utilize the compiled matrix
multiplication utility in Matlab on a squared single summation. One can avoid loops by computing
vectorized sums so that the element-wise operations are made in one
stroke using pre-compiled functions. In addition, by pre-dimensioning
the matrices, one can reduce the computational time even further. A suggested sample
Matlab code can be found in Appendix B. Using this method, the spectra presented here can be computed in a
matter of minutes on a 1.2\,GHz dual processor laptop. Note
that the blocks should have the same record length, T, in physical
units, so the number of samples per block might vary. Keeping T
constant insures that the window function convolving the spectrum is
constant. Further, it is also important to remember to subtract the
(residence time weighted) mean velocity of each particular block,
not that of the full time series. Obviously care should be taken to
insure that blocks with wildly different values of the number of
realization are eliminated from the analysis.

\begin{acknowledgements}
The authors are grateful to Dr. Holger Nobach, Max-Planck Institut f\"{u}r Dynamik und Selbstorganisation for providing the cylinder wake data. The authors also gratefully acknowledge many helpful discussions with Dr. T. Gunnar Johansson of Chalmers Technical University. This work formed a portion of the Ph.D. dissertation of CMV, who was supported by the Danish Research Council, DSF, under grant number 2104-04-0020. WKG acknowledges the support of Vetenskapsr{\aa}det (The Swedish Research Council). This work was carried out in part while WKG was an Otto M{\o}nsted Visiting Professor at the Danish Technical University in 2007.
\end{acknowledgements}

\newpage
\section*{Appendix A: High-velocity and directional bias correction}

It is clear that the method of residence-time weighted data processing recapitulated above provides the correct statistical results, since it was shown to be equivalent to conventional time averages. Thus the concept of bias does not enter at all. However, the method is only correct for uniformly seeded flows of constant, uniform density. It was also assumed that no other sources of bias existed, e.g. directional bias due to the finite fringe number effect or electronic biasing.

The concept of bias in burst-type LDA measurement and methods of bias correction were introduced in 1973, when it became clear that the results of direct arithmetic averaging contained errors, so-called bias errors, especially for measurements in high intensity turbulence. One of the first treatments on the problem was given by McLaughlin and Tiederman~\cite{McLaughlin}, who defined the problem and proposed a method of correction based on the weighting of the measured data, $u_0(t_i)$, with the inverse of the numerical value of the measured velocity component, i.e., with the factors $u_0(t_i)^{-1}$. The algorithms for the computation of mean and mean-square values by this method, from here on termed the one-dimensional or 1-D correction, are:
\begin{equation}
\overline{u(\mathbf{x_0})}_{1-D}=\frac{\sum_i u_0(t_i) |u_0(t_i)|^{-1}}{\sum_i |u_0(t_i)|^{-1}}
\end{equation}
and
\begin{equation}
\overline{[u'(\mathbf{x_0})]^2}_{1-D}=\frac{\sum_i [u_0(t_i) - \overline{u}(\mathbf{x_0})]^2 |u_0(t_i)|^{-1}}{\sum_i |u_0(t_i)|^{-1}}
\end{equation}

\begin{figure}
  \center{\includegraphics[width=0.75\textwidth]{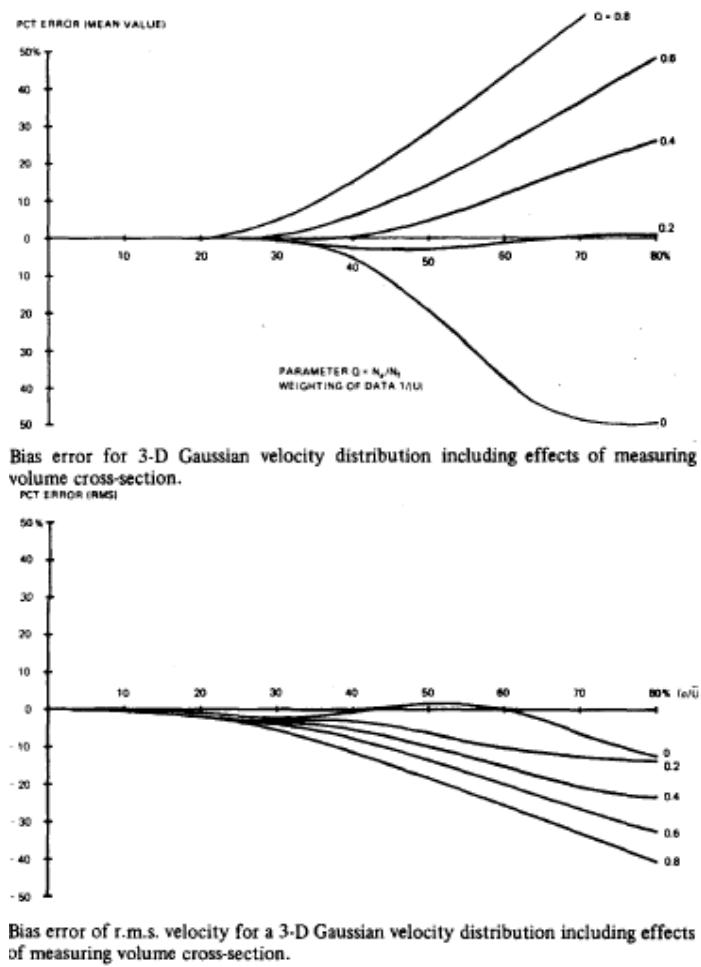}}
\caption{Errors in computed mean and mean-square velocity as a function of turbulence intensity.} \label{fig:Fig45}
\end{figure}

McLaughlin and Tiederman investigated the accuracy of the 1-D correction by numerical simulation of certain 2-dimensional flow cases. However, they did not consider three dimensional cases and did not consider the directional bias due to the finite fringe number effect\footnote{Sometimes called ``processor directional bias''. Directional bias was important when LDA counters were used because the counter needed a certain minimum number of fringes to accept a data point, and that certain flow directions might not produce a signal with enough fringes. But the directional bias disappears when enough frequency shift is used. Later processors based on FFT analysis of the signal do not suffer processor directional bias at all due to the processing method.}. Buchhave~\cite{Buchhave1976} investigated the accuracy of l-D correction data by simulating data from isotropic turbulence, and included the directional bias. It was shown that the l-D correction over-compensates for the velocity bias and leads to large errors for turbulence intensities above 15-20\%. However, the angular bias, which occurs for many types of LDA counters used up to this time, pulls in the other direction, and under optimal conditions quite good results may be obtained in certain situations even for turbulence intensities in the order of 50\%. Figure \ref{fig:Fig45} from~\cite{Buchhave1976} shows the errors in computed mean and mean-square values as a function turbulence intensity of an isotropic, Gaussian turbulence superimposed on a constant mean velocity.

The problem of bias was discussed for the first time at the 1974 Purdue meeting (see e.g.~\cite{Durst1974}). The theory was further developed by~\cite{7} along the lines presented earlier in the current work. Hosel and Rodi~\cite{Hoseletal1977} also developed the theory from considerations on the average data sampling rate for varying velocities and provided some experimental results to support the theory. Later Erdman and Gellert~\cite{ErdmanGellert1976} studied the correlation between velocity and particle arrival rate and showed results which substantiate the ideas about proportionality between data rate and velocity in one-dimensional flows.

Some investigators, however, report a weaker correlation between velocity and data-rate than expected from the preceding theory (see e.g.~\cite{SmithMeadows1974}). In these cases electronic noise may have caused secondary effects, since these measurements were made at rather high velocities.

More recent measurements by Karpuk and Tiederman~\cite{KarpukTiederman1976} and Quigley and Tiederman~\cite{QuigleyTiederman1977} in the viscous sublayer in pipe flow show good agreement between mean and rms values computed by the l-D correction method and hot-wire data. The measurements were made with an optical system especially designed to give a probe volume with small spatial dimensions in the direction of the mean velocity gradient. The authors note that the data rate did not seem to be correlated with direction, but unfortunately no details are given on the ratio of the number of fringes needed for operation of the signal processor to the total number available in a burst.

Finally, concerning velocity bias it may be mentioned (as already pointed out by McLaughlin and Tiederman) that in simultaneous measurement of all three velocity components (or in the case of two-dimensional measurements in flows in which the fluctuations in the third direction are negligible) it is of course possible to compute the magnitude of the velocity $\mathbf{u_i}$ and assign a weighting factor proportional to $|\mathbf{u_i}|^{-1}$ to each sample set ($u_i,v_i,w_i$), and even to correct for directional bias based on the knowledge of the direction of $\mathbf{u}$. Such corrections might be carried out on stored data-points after the measurement, but it appears that this method has not yet been tried in actual measurements\footnote{This type of directional bias is caused by the variation of the geometrical cross section as viewed from different flow directions when the measuring volume is not spherical (e.g. if it is an elongated ellipsoid). Here we should stress, what few realize when discussing this subject, that residence time weighting also removes this ``measurement volume directional bias'' completely. This is because the product of the geometrical cross section and the average path length (which produces the residence time in the measurement) is a constant equal to the volume of the measuring volume. Thus, no matter what the geometrical shape of the measuring volume is, the directional bias is non-existent. (when the measuring volume cross section is small the corresponding residence times are long). There is an additional noise due to the fluctuations in the length of the residence time at different paths through the measuring volume. When calculating this effect by comparing the direct spectral estimator for one of the measured records with measured residence times and constant residence times, the added noise is found to be only about 3\,dB.}.

The question of whether other factors influence the data-rate is still largely unresolved. Tiederman~\cite{Tiederman1977} raised the question of whether differences in signal-to-noise ratio of fast and slow bursts might influence the data rate. Other physical effects that cause correlation between data rate and flow velocity include density variations caused by pressure or temperature fluctuations, mixing of fluids with different particle concentration, and chemical reactions. Asalor and Whitelaw~\cite{AsalorWhitelaw1976} derived expressions for the correlation between combustion induced temperature, pressure and concentration fluctuations, and computed the data rate based on assumptions about the velocity-temperature and velocity-pressure correlations in a diffusion flame. From this analysis and subsequent measurements the authors concluded that in this particular flow the bias effects due to velocity fluctuations confirmed the velocity-data rate correlations expected from the residence time analysis and e.g. Maclaughlin and Tiederman's assumptions. Velocity-pressure and velocity-temperature correlation effects were found to be negligible.

George~\cite{7} discusses briefly the extension of the arguments leading to the residence-time weighting to flows with density fluctuations. No previous attempts seem to have been made on weighting or bias-correction in LDA measurements of correlation functions or spectra.

The slotted-time-lag autocovariance method developed by Gaster and Roberts~\cite{14} was simultaneously being developed by Mayo and others (see e.g.~\cite{SmithMeadows1974,Mayo1974,Scott1974}). Curiously no one seemed to notice that the time-slots themselves re-introduced  the aliasing that was to have been eliminated by the random sampling. Measurements on the turbulence of a free jet were reported by Smith and Meadows~\cite{SmithMeadows1974}. The basic feasibility of measurements of turbulence power spectra with burst-type LDAs was proven. Later also other measurements were reported (Mayo et al.~\cite{MayoShayRiter1974} and Bouis et al.~\cite{BouisGourotPfeiffer1977}). However none of these report any attempt to consider weighting of the data along the lines discussed above or to correct for biasing. Wang~\cite{Wang1976} and Asher, Scott and Wang~\cite{AsherScottWang1974} discuss various sources of error and noise in LDA-counter measurements of power spectra and conclude that the quantizing of the output from the LDA-counter due to the finite resolution of the counter itself is the greatest source of error. However, these reports did not consider the ``apparent turbulence'' caused by the finite dimensions of the measuring volume in the presence of gradients within the volume as described in~\cite{3}, Chapter 4.5, nor did it consider the biasing effects introduced by using uncorrected data. It should be apparent that bias effects will modify the computed spectra as well as mean and mean square values, and that bias correction methods should be applied to spectral measurements as well. The theoretical considerations presented earlier indicate that the phenomenon of bias need not exist in the sense that it had been considered here and results only from incorrect signal processing.

\newpage
\section*{Appendix B: Proposed Matlab-algorithm}

\begin{verbatim}
%% Load velocities, arrival times and transit/residence times
load u; load t; load tt;
%% Make sure vectors are arranged as 1 by N vectors
u=u(:)'; t=t(:)'; tt=tt(:)';

M = 600;                % Number of frequencies (arbitrary)
f = logspace(1,4,M);    % Define frequencies (arbitrary)
T_b = 0.1;              % Block record length [s]
bl = floor(t(end)/T_b); % Number of blocks

%% Extract index of first element in each block
blocklength = linspace(1,bl+1,bl+1)*T_b; %% Set block record length
ii(1) = 1;    % Index of first element in block
for k=1:bl+1
    % Determine ii - start index for each block
    [rte(k),ii(k+1,1)] = min(abs(t - blocklength(k)));
    % RTW average of u for each block
    umean(k) = sum(u(ii(k):ii(k+1)-1).*tt(ii(k):ii(k+1)-1))/...
                (sum(tt(ii(k):ii(k+1)-1)));
    % Obtain velocity fluctuations
    u(ii(k):ii(k+1)-1) = u(ii(k):ii(k+1)-1)-umean(k);
end
%% Element-wise multiplication of the velocities and transit times
utt = u.*tt;
N = max(diff(ii))+1; % Maximum number of samples in a block

%% Pre-dimension and allocate matrices to reduce
%% need for memory handling resources
uf = zeros(M,bl);
f_tot = repmat(f',1,N);
T=zeros(1,bl);
sumtt=zeros(1,bl);

%% Compute variables for each block separately
for k = 1:bl
    % Fourier transform of velocity
    uf(:,k) = exp(-sqrt(-1)*2*pi*repmat(t(ii(k):ii(k+1)-1),length(f),1)...
        .*f_tot(:,1:ii(k+1)-ii(k))) * utt(ii(k):ii(k+1)-1)';
    % Precise block record length
    T(k) = (t(ii(k+1)-1)-t(ii(k)));
    % Sum of residence/transit times
    sumtt(k) = sum(tt(ii(k):ii(k+1)-1));
end

%% Compute and average over all blocks
Sblocks = repmat(T./sumtt.^2,M,1).*abs(uf.*conj(uf)); Sblocks(isnan(Sblocks))=0;
S = mean(Sblocks,2); loglog(f,S);
\end{verbatim}

\newpage
\section*{Appendix C: Errors from omitting residence time weighting}

As was described by~\cite{4}, the effect of computing the arithmetic statistical moments instead of the residence time weighted ones can be evaluated by performing a Taylor expansion. For both the mean value and the variance (which the power spectrum is directly linked to), a Taylor expansion has been performed about the state of zero turbulence intensity. It is assumed that all particles pass through the center of the measurement volume and traverse undisturbed. Further, only a one-dimensional flow is considered, say $\tilde{u} = U+u$, where $U$ is the mean and $u_n$ the departure of the $n^{th}$ realization from it.

It is important to note that, from the expansions, it is apparent that large values of higher order moments can affect the accuracy significantly. This is usually not the case at the jet centerline, where the mean velocity is the largest and the velocity variance is the main higher order term. But as one moves further and further off the centerline, the mean velocity decreases and the impact of the higher moments (skewness, kurtosis, etc.) increase, resulting in larger errors.

First, let's consider the residence time weighted mean velocity:
\begin{equation}
U_{0,T}^{B} = \frac{\sum_{n=0}^{N-1}\tilde{u}_n \Delta t_n}{\sum_{n=0}^{N-1}\Delta t_n}
\end{equation}

The above assumptions imply the following approximate form for the sum of residence times:
\begin{equation}
\sum_{n=0}^{N-1}\Delta t_n \approx \sum_{n=0}^{N-1} \frac{d}{U+u_n} = \frac{d}{U}\sum_{n=0}^{N-1} \frac{1}{1+u_n/U}
\end{equation}
where $d$ is the measuring volume length. Expanding the denominator yields
\begin{equation}
\sum_{n=0}^{N-1}\Delta t_n \approx \frac{d}{U}\sum_{n=0}^{N-1} \left \{ 1 - \frac{u_n}{U} + \frac{u_n^2}{U^2} - \frac{u_n^3}{U^3} + \frac{u_n^4}{U^4} - \ldots \right \}
\end{equation}
\begin{equation}
\approx N\frac{d}{U} - 0 + \frac{d}{U}\sum_{n=0}^{N-1}\frac{u_n^2}{U^2} - \frac{d}{U}\sum_{n=0}^{N-1}\frac{u_n^3}{U^3} + \frac{d}{U}\sum_{n=0}^{N-1}\frac{u_n^4}{U^4} \approx N\frac{d}{U} \left \{ 1 + \frac{\overline{u^2}}{U^2} - \frac{\overline{u^3}}{U^3} + \frac{\overline{u^4}}{U^4}\right \}
\end{equation}

Similarly, the Taylor expansion of the residence time weighted mean velocity can be expanded to second order in $\overline{u^2}/U^2$ to obtain:
\begin{equation}
U_{0,T}^{B} \approx \frac{Nd}{N\frac{d}{U}\left \{ 1 + \frac{\overline{u^2}}{U^2} \right \}} \approx U \left \{ 1 - \frac{\overline{u^2}}{U^2} + \frac{\overline{u^3}}{U^3} - \frac{\overline{u^4}}{U^4} + \ldots \right \}
\end{equation}
Thus, the deviation of the arithmetic mean to the residence time weighted one is proportional to the variance and inversely proportional to the mean velocity.
\begin{equation}
\frac{\Delta U_{0,T}^{B}}{U} = \frac{U_{0,T}^{B} - U}{U} \approx -\frac{\overline{u^2}}{U^2}
\end{equation}
It is clear from the above results, which were confirmed for the data sets presented in section~\ref{sec:4} in~\cite{4}, that the turbulence intensity is crucial for low bias errors in the mean value. \\

The same procedure can be applied to the higher moments. Consider the second moment (which is vital in estimating the power spectrum), where
\begin{equation}
\overline{u^2}_{0,T}^{B} \approx \frac{d\sum_{n=0}^{N-1}(U+u_n)}{N\frac{d}{U}\left \{ 1 + \frac{\overline{u^2}}{U^2} \right \}} \approx  \frac{NdU}{N\frac{d}{U}\left \{ 1 + \frac{\overline{u^2}}{U^2} \right \}} \approx U^2 \left \{ 1 - \frac{\overline{u^2}}{U^2} + \left ( \frac{\overline{u^2}}{U^2} \right )^2 + \ldots \right \}
\end{equation}
which yields
\begin{equation}
\frac{\Delta \overline{u^2}}{U^2} = \frac{\overline{u^2}^B_{0,T} - \overline{u^2}}{U^2} \approx 1-2\frac{\overline{u^2}}{U^2}+\left ( \frac{\overline{u^2}}{U^2} \right )^2
\end{equation}
and so on. \\

\end{document}